\documentclass[apj]{emulateapj}
\usepackage[dvips]{epsfig,color}
\usepackage{amsmath}

\newcommand{\eg}{{\it e.g.,}}

\newcommand{\etal}{{\it et al.}}

\DeclareMathAlphabet{\mathsfsl}{OT1}{cmss}{m}{sl}

\newcommand{\bs}[1]{\boldsymbol{#1}}
\newcommand{\dif}{\mathrm{d}}
\newcommand{\avps}{\ensuremath{\langle v_p^2 \rangle}}
\newcommand{\qvec}{\ensuremath{\frac{\rho}{2} 
\langle v_p^2 \boldsymbol{v_p} \rangle}}
\newcommand{\vnab}{\ensuremath{\boldsymbol{\nabla}}}
\newcommand{\vvz}{\ensuremath{\boldsymbol{v_0}}}

\newcommand{\bcdot}{\ensuremath{\boldsymbol{\cdot}}}

\newcommand{\kb}{\ensuremath{k_{\rm B}}}

\newcommand{\cs}{\ensuremath{\chi_r^2}}
\newcommand{\vrms}{\ensuremath{v_{\rm rms}}}
\newcommand{\cslb}{\ensuremath{\chi_{\rm LB}^2}}
\newcommand{\csmb}{\ensuremath{\chi_{\rm MB}^2}}
\newcommand{\cslbi}{\ensuremath{\chi_{\rm LB,iso}^2}}
\newcommand{\cslba}{\ensuremath{\chi_{\rm LB,aniso}^2}}
\newcommand{\csmbi}{\ensuremath{\chi_{\rm MB,iso}^2}}
\newcommand{\csmba}{\ensuremath{\chi_{\rm MB,aniso}^2}}

\shorttitle{Entropy Production}
\shortauthors{Barnes \& Egerer}

\begin{document}

\title{Entropy Production in Collisionless Systems. III. 
Results from Simulations}
\author{Eric I. Barnes}
\affil{Department of Physics, University of Wisconsin --- La Crosse,
La Crosse, WI 54601}
\email{barnes.eric@uwlax.edu}
\author{Colin P. Egerer}
\affil{Department of Physics, University of Wisconsin --- La Crosse,
La Crosse, WI 54601}
\email{egerer.coli@uwlax.edu}

\begin{abstract}

The equilibria formed by the self-gravitating, collisionless collapse
of simple initial conditions have been investigated for decades.  We
present the results of our attempts to describe the equilibria formed
in $N$-body simulations using thermodynamically-motivated models.
Previous work has suggested that it is possible to define distribution
functions for such systems that describe maximum entropy states.
These distribution functions are used to create radial density and
velocity distributions for comparison to those from simulations.  A
wide variety of $N$-body code conditions are used to reduce the chance
that results are biased by numerical issues.  We find that a subset of
initial conditions studied lead to equilibria that can be accurately
described by these models, and that direct calculation of the entropy
shows maximum values being achieved.

\end{abstract}


\section{Introduction}\label{intro}

Over the past several decades, numerous investigations of
collisionless, self-gravitating systems have been undertaken.  From
early focus on the formation and evolution of elliptical galaxies
\citep[e.g.,][]{va82}, to cosmological simulations of dark matter
structure formation \citep[e.g.,][]{nfw96,moore99,setal05}, these
works have taken advantage of ever-increasing levels of computing
power.  Simulations with higher and higher resolutions (numbers of
particles per unit volume) are constantly being performed.
Our goal is not to attempt to replicate these state-of-the-art
simulations, but rather to use more modest simulations to investigate
some basic questions.  The systems that we simulate are not direct
analogues of putative dark matter halos nor elliptical galaxies, but
they do share the fundamental physical conditions of self-gravitation
and collisionless evolution.  As many previous works have shown
evidence of ``universal'' behaviors [such as radial density
\citep[e.g.,][]{nfw97,n04} and power-law ``phase space''
$\rho(r)/\sigma^3(r)$ \citep{tn01} profiles], an interesting
possibility is that a basic physical mechanism may underlie the
formation of these self-gravitating equilibria.

Earlier work \citep{bw11,bw12} presents descriptions of distribution
functions of collisionless, self-gravitating systems that represent
maximum entropy states.  These results follow the seminal work of
\citet{lb67}, where it is argued that a fourth statistical family is
appropriate for describing the phase-space evolution of these kinds of
systems.  In a nutshell, the familiar Maxwell-Boltzmann statistics
describe systems in which particles are distinguishable and do not
obey a phase-space exclusion principle -- multiple particles can
occupy a very small region of phase-space.  On the other hand, the
Lynden-Bell statistical family is appropriate for systems of
distinguishable particles that follow an exclusion principle -- a
classical version of Fermi-Dirac statistics.  By relaxing the
requirement of large phase-space occupation numbers, \citet{bw12} show
that finite-mass, maximum entropy states exist for both Lynden-Bell
and Maxwell-Boltzmann statistical families.  The major goal of this
work is to test whether or not any simulated system will relax to such
a state.  It is certain that these models will fail for sufficiently
strong collapses, as the assumed velocity isotropy underlying the
models has to disappear as the radial orbit instability begins to
become important \citep[\eg][]{ma85,blw09}.  As such, we also aim to
identify ranges of conditions that can result in maximum entropy
states.  A final goal is to monitor the behavior of entropy in
simulations to see if a maximum value is reached.

We have created suites of simulations to test the usefulness of the
Lynden-Bell and Maxwell-Boltzmann families of models.  The
publicly-available GADGET code \citep{s05} has been employed to evolve
simulations of collisonless systems comprised of $N=10^5$ and $N=10^6$
particles.  These simulations approximate collisionless conditions by
utilizing softened interactions.  As a point of comparison, we have
also analyzed collisional systems with $N=2^{17} \approx 10^5$
particles using a version of NBODY-6, enhanced with a Graphics
Processing Unit (GPU) \citep{na12}.  This code uses direct Newtonian
particle-particle interactions, but the large number of particles
guarantees that two-body relaxation processes occur over timescales
thousands of times longer \citep[][Ch. 4]{bt87} than any gravitational
potential (or ``violent'') relaxation processes, which occur over a
few initial crossing times $T$.

For any given simulation, we analyze spherically symmetric density and
velocity distributions by counting particles in spherical bins
centered on the system center-of-mass.  With estimates of the various
uncertainties, these radial profiles are then used as the data to be
matched by the Lynden-Bell and Maxwell-Boltzmann models.  Chi-squared
minimizations are used to indicate the appropriateness of each model.
We will not insist that models with low chi-squared values are the
only descriptions of these simulations, merely that such a model is
consistent with the data.  For simulations that are well-described by
the thermodynamic models, we also investigate the behavior of entropy
during its evolution.

We begin by describing simulation initial conditions, evolution code
details, and analysis techniques in section~\ref{sims}.  Two methods for
describing entropy behavior in these simulations are presented in
Section~\ref{entropy}.  Section~\ref{fitting} contains the findings
inferred from minimizations for selected simulations, while
section~\ref{entprod} outlines the entropy behaviors seen in the
simulations.  We summarize in Section~\ref{conclude}.

\section{Simulation Details}\label{sims}

\subsection{Initial Conditions}

For all simulations discussed here, particles are assumed to be
identical and system mass, system radius and Newton's gravitational
constant are set to unity ($m_{\rm sys}=R=G=1$).  Each system is
composed of $N$ particles.  Following \citet{va82}, particle positions
are chosen according to two different schemes.  

In ``single'' simulations, initial particle locations are randomly
chosen within the system according to a specified density distribution
-- cuspy ($\rho \propto 1/r$) or Gaussian ($\rho \propto \exp{-r^2}$).
A simple rejection scheme is used to generate the distributions.
The system center-of-mass must coincide with the center of the
spherical boundary to better than $1/\sqrt{N}$ to be acceptable.

For ``clumpy'' simulations, centers-of-mass locations are chosen for
small clumps of particles according to the above density
distributions, and then particles are uniformly distributed within a
clump.  The numbers of particles in each clump are chosen from a
Salpeter distribution -- the probability of generating a clump with
$N_{\rm clump}$ particles is proportional to $N_{\rm
clump}^{-\alpha}$, where $\alpha=2/3$ for this work.  As we are
not investigating any specific physical situation with these
simulations, the details of this distribution (for example, variations
to the adopted $\alpha$) have not been scrutinized.  The Salpeter form
has been adopted based on its simplicity.  We demand that the sum of
the volumes of all the clumps equals that of a sphere with radius
$R=1$.  Individual clump radii are chosen proportional to the fraction
of the total mass they contain, $r_{\rm clump} \propto (N_{\rm
clump}/N)^{1/3}$.  Clumps can overlap one another, leading these
initial conditions to have regions of higher-than-average density as
well as nearly empty regions where clumps fail to overlap.

Particle velocities are given random orientations, guaranteeing
initial velocity isotropy.  Speeds are chosen by adopting an initial
virial ratio $Q_0=2K_0/|W_0|$ that links a system's initial kinetic
energy $K_0$ to its initial potential energy $W_0$.  Once particle
positions have been selected, the virial ratio is used to define a
scale speed.  For single simulations, this is the speed given to every
particle in the system.  Clumpy simulations distribute speeds in a
more complicated manner.  We define hot-clumpy systems to be ones in
which the clump centers-of-mass have zero initial velocities --
particles in clumps are given random velocity directions with equal
speeds.  Cold-clumpy systems are composed of clumps in which all of
the particles move with the clump center-of-mass velocity.  The
centers-of-mass velocities are randomly oriented but have the same
magnitudes.  As an intermediate case, warm-clumpy simulations split
the kinetic energy equally between individual particle motion and
clump center-of-mass motion.  Independent of the specifics of the
setup, the systems are not initially in mechanical equilibrium, even
though they may be in virial equilibrium (if $Q_0=1$ is adopted).  

\subsection{Evolution Code Details}

As mentioned in the introduction, this work utilizes two very
different codes for evolving initial conditions.  Our aim is to be
able to identify any numerical effects due to particle number,
softening parameters, and/or code specifics.  The GPU-enhanced
NBODY-6 code has an architecture designed for investigating globular
cluster dynamics.  GADGET has been designed to perform cosmological
simulations of structure formation.  Finding agreement between our
predictions and the results of both types of simulations will
strengthen the assertion that our analytical picture is relevant to
collisionless systems and is not biased by numerical issues.

GADGET is a versatile tree-code that incorporates softened forces.  In
general, we have adopted the standard parameters for GADGET.  However,
we do not evolve in an expanding universe, and we adopt different
softening lengths, depending on the situation.  For $N=10^5$
simulations, we adopt a softening length $\epsilon=10^{-4}$, about 100
times smaller than the ``optimal'' softening length value described in
\citet{p03}.  Test runs with softening lengths between our adopted
value and the optimal value have resulted in only minor differences in
density profile shape.  Smaller values lead to integration times that
we judged to be unacceptably long, so our value is as small as
possible while keeping the wall-clock time manageable.  For $N=10^6$
simulations, we adopt the optimal softening length $\epsilon=4\times
10^{-3}$.  Again, testing with smaller softening lengths indicates
that density profiles are largely unaffected, but integration times
significantly lengthen.

We have performed three types of GADGET simulations.  For $N=10^5$, we
have varied $Q_0$ between 0.1 and 1.0 for both types of initial
density distributions and for the three different velocity assignments
when starting with clumpy initial conditions.  We take these as the
standard set of simulations that provide zeroth-order tests of our
models.  As the initial conditions are based on random distributions
of particles, we have also performed ensemble simulations of a subset
of these initial conditions.  Five independent realizations of initial
conditions with $0.7 \le Q_0 \le 1.0$ and both initial density
distributions have been evolved.  For clumpy simulations, only
ensembles with hot conditions were created.  Section~\ref{fitting}
discusses the justification for the ranges and specific values used.
Averages of these five realizations have been used to validate the
results of the standard simulations.  The third type of GADGET
simulation involves $N=10^6$ particles.  Both initial density
distributions with $0.7 \le Q_0 \le 1.0$ have been investigated for
single and clumpy systems.  Like the ensemble simulations, these
additional results provide an estimate of the robustness of the
results based on the $N=10^5$ simulations.

The GPU-NBODY-6 simulations evolve single and clumpy systems with $N
\ga 10^5$, $0.7 \le Q_0 \le 1.0$, and both initial density profiles.  This
code does not utilize softened forces, so it is a collisional code.
However, the large number of particles provides a reasonable basis for
the assumption that two-body effects (\eg\ ejection) have relatively
minor impact on the simulations.  For example, no particles gain
escape speeds during any of our simulations due to two-body
collisions.  Unfortunately, simulations with larger particle numbers
could not be completed with the current hardware available to the
authors.

\subsection{Analysis}\label{analysis}

Independent of the evolution code, each simulation extends for at
least 20 initial crossing times.  We have observed that this is
generally a sufficient period for a system to reach a mechanical and
virial equilibrium state.  For clumpy initial conditions,
individual clumps have ``dissolved'' by 10 initial crossing times, at
the latest.  By construction, clumps in the hot-clumpy simulations
begin to disperse almost immediately.  Any simulation with a longer
evolution will be noted in what follows.  We infer mechanical
equilibrium by observing that the radius of the inner-most 90\% of the
particles stops varying and that the average velocities of the
particles are zero over the radial range of the inner-most 90\% of
particles.  Simulations with $Q_0 \ga 0.3$ typically reach these
conditions in less than 20 initial crossing times.  Systems are
decomposed into spherical, 1-percent mass shells at each output
timestep.  Particles in these shells provide density and velocity
statistics (averages, rms values, and variances).  Each shell is
broken into three sub-shells which provide ranges for the density and
velocity values that are used to estimate uncertainties (maximum minus
minimum).  Systemic axis ratios, phase-space occupation values, and
entropy production rates (see section~\ref{entropy}) are also
calculated at each output timestep.

With the radial density $\rho$ and rms speed \vrms\ distributions
formed, we next compare to several model distributions.  Our focus is
on the comparison to the Lynden-Bell and Maxwell-Boltzmann models, but
we also include two common analytical models, Plummer and de
Vaucouleurs.  The Plummer models we use have a density distribution
given by \citep{p11,bt87},
\begin{equation} \rho=\frac{\rho_{0}}{\left(1+\frac{1}{3}
\left(\frac{r}{r_{0}}\right)^{2}\right)^{5/2}}, 
\end{equation} where
$\rho_{0}$ is a scaling density and $r_{0}$ is a scaling radius.  This
density profile is relatively constant in the core of the system and
declines rapidly ($\propto r^{-5}$) near the outer edge.  The de
Vaucouleurs profile \citep{dv48} has been used to fit the light
profiles of elliptical galaxies, and is a simple example of a broken
power-law distribution \citep[a cousin to commonly-discussed,
cosmologically-motivated profiles such as,][]{nfw96,moore99}.  It is
well-established that the type of simulations discussed here result in
structures with outer density behavior that matches the de Vaucouleurs
profile when $Q_0 \la 0.2$ \citep{va82,sl12}, so it provides a good
benchmark for our standard simulations.  The de Vaucouleurs density
profile is given by,
\begin{equation}
\rho=\rho_{0}\left(\frac{r}{r_{0}}\right)^{-\delta}
\left(1+\frac{r}{r_{0}}\right)^{\delta-4},
\end{equation}
where $\delta=\frac{1}{2}$, and $\rho_{0}$ and $r_{0}$ are again
a scaling density and radius, respectively. A de Vaucouleurs density
profile has a central cusp, in contrast to the central density core
behavior of the Plummer model.

As our models and simulated systems all have finite masses, we demand
that their density and \vrms\ values match at the half-mass
radius.  With the connection between model and simulated data values
fixed, we use the reduced chi-squared statistic as the
figure of merit for our fits,
\begin{equation}
\cs=\frac{1}{N_{\rm data}}\sum_{i=1}^{N_{\rm
data}}\frac{(M_i-D_i)^{2}}{\Delta_i^{2}},
\end{equation}
where $N_{\rm data}=100$ for our 1\% mass shells, $D_i$ is a mass
shell density or \vrms\ value from a simulation, $M_i$ is a model
value corresponding to the same radial location, and $\Delta_i$ is an
uncertainty estimate for the simulation value (as described in
the first paragraph of this section).  One should expect, if
the uncertainty estimates are appropriate, that a good model fit to
the data produces $\cs \approx 1$.  

The Plummer and de Vaucouleurs density models have no free parameters,
so their \cs\ values are determined upon matching to simulation values
at the half-mass radius.  For Lynden-Bell and Maxwell-Boltzmann
models, a single parameter ($\nu_{\rm LB}$ or $\nu_{\rm MB}$)
determines the shape of the density profile, and hence the match to
the simulation.  Density profiles are determined by iteratively
solving the Poisson equation in straightforward fashion \citep[][Sec.\
4.4.2]{bt87}.  The best-fit value of $\nu$ is determined using an
amoeba \cs\ minimization \citep{press94}.  We have also performed
Markov Chain Monte Carlo minimizations to corroborate the amoeba
results.

To determine model \vrms\ distributions, we solve the Jeans equation
for a given density profile.  This approach demands a choice be made
regarding the radial behavior of the velocity anisotropy $\beta(r)$.
We utilize two \vrms\ profiles: one assuming velocity isotropy
$\beta(r)=0$ and one adopting $\beta(r)$ from the simulated system.
The Lynden-Bell and Maxwell-Boltzmann models are derived assuming
velocity isotropy, so their density profiles are consistent only with
the $\beta(r)=0$ \vrms\ profiles.  In the absence of a
distribution function that incorporates the mild tangential velocity
anisotropy present in our systems, we assume that the density derived
from such a function should be well-approximated by the density
resulting from the isotropic version of the distribution function.
While not an exact description of the situation in our work, this
assumption is consistent with the behavior of the aniostropic Plummer
model described in \citet{m85}.  We choose the $\beta(r)$ to be used
in the Jeans equation to be a smoothed version of the velocity
anisotropy present in the simulation.  Specifically, a fourth-order
polynomial fit to a simulation anisotropy profile is created.  The
parameters describing the Lynden-Bell and Maxwell-Boltzmann models are
not allowed to vary during comparison to the \vrms\ distribution, so
the velocity \cs\ calculation for all models is straightforward once
the model and simulated \vrms\ values are matched at the half-mass
radius.

\section{Entropy Behavior}\label{entropy}

\subsection{Microscopic Picture}\label{micro}

The basis of the \citet{bw11,bw12} work is the phase-space counting
approach outlined in \citet{lb67}.  Here, we present a brief summary
of the chief ideas necessary for defining entropy from this
microscopic viewpoint.  Phase-space is imagined to be sub-divided into
two lattices; an array of nearly infinitesimal micro-cells (each with
volume $\varpi$) that are arranged into collections of macro-cells.
Macro-cells contain $\nu$ micro-cells, and this value serves as the
control parameter for the models.  Micro-cell occupation defines a
fine-grained distribution function, while macro-cell occupation
defines a coarse-grained distribution function that can be realized
through simulations.  The occupation of micro-cells determines the
statistical properties of the system.  Maxwell-Boltzmann statistics
arise when multiple distinguishable particles can occupy a micro-cell.
Lynden-Bell statistics describe situations in which a classical
exclusion principle disallows multiple particles occupying a single
micro-cell.  For collisionless systems like the ones we investigate
here, the fine-grained distribution function is a constant of motion.
As such, one expects the Lynden-Bell statistical family to be most
appropriate because the fine-grained distribution function values
cannot be increased through multiple occupancy.

With these ideas, one can count the number of energy states available
to a system, and hence, define the entropy.  The \citet{lb67} work
relies on the Stirling approximation to simplify entropy calculations,
but \citet{bw12} relax this assumption and allow macro-cell occupation
numbers to be small enough that the Stirling approximation fails.  The
end result is that the Lynden-Bell entropy can be given as
\citep[Equation 13 in][]{bw12},
\begin{eqnarray}\label{slb}
\lefteqn{ S_{\rm LB} = S_{\rm LB,0} - } \nonumber \\
& & \kb \sum_{i=1}^M \left[
(n_i+1/2)\ln{(n_i+1)} + \right. \nonumber \\
& & \left. (\nu -n_i +1/2)\ln{(\nu-n_i+1)} +
\lambda_{0,n_i} + \lambda_{0,(\nu-n_i)} \right],
\end{eqnarray}
where $S_{\rm LB,0} = \kb [ N\ln{N} - N + M(\nu \ln{\nu} -
\ln{2\pi})]$, $N$ is the number of phase-space elements/particles, and
$M$ is the total number of macro-cells.  The $\lambda$ function arises
from the approximation,
\begin{equation}\label{approx}
\ln{x!}=(x+\frac{1}{2})\ln{(x+1)} - x + \frac{\ln{2\pi}}{2} +
\lambda_{0,x}
\end{equation}
where 
\begin{equation}
\lambda_{0,x}=-\frac{(x^2 + 2x + \frac{287}{288})}{(x^2 +
\frac{25}{12}x + \frac{13}{12})}.
\end{equation}
A similar expression can be found for the Maxwell-Boltzmann entropy
\citep[see Equation A2 in][]{bw12}.  Adopting these modifications
leads to the possibility of finite-mass and energy systems that belong
to the Maxwell-Boltzmann statistical family, in contrast to the
findings of \citet{lb67}.  Likewise, the discussion in \citet[][\S
4.7.1]{bt87} regarding the impossibility of maximizing entropy becomes
invalid, as the simple $f\ln{f}$ term in the entropy calculation is
now modified.  Our Lynden-Bell expression does not change the overall
character of the associated distribution function presented in
\citet{lb67}; it remains finite-mass and energy and closely resembles
a Fermi-Dirac distribution.

We have calculated $S_{\rm LB}$ using the results of GADGET $N=10^6$
simulations.  At every timestep, positions and velocities are used to
assign each particle to a macro-cell, giving $n_i$.  A fixed value of
$\nu_{\rm LB}=10^4$ has been chosen for these calculations, as that is
always greater than the maximum $n_i$ value in these simulations.
Tests varying $\nu_{\rm LB}$ show that the ``zero point'' of $S_{\rm
LB}$ is affected much more strongly than its time-dependent behavior .
Results of these calculations are discussed in Section~\ref{entprod}.

\subsection{Macroscopic Picture}\label{macro}

As a complement to the microscopic approach, we follow the discussion
of thermal non-equilibrium situations given by \citet{dgm84}.  The
behavior of self-gravitating systems composed of a large number of
massive particles can be described using equations that represent
macroscopic conservation laws.  Expressions of the conservation of
energy can be manipulated and combined with the first law of
thermodynamics to provide insight into the behavior of entropy.  In
particular, it is useful to write the entropy production rate of a
system as,
\begin{equation}\label{s4}
\frac{\partial S^{(c)}}{\partial t} = \int_{\rm V} \sigma \, \dif^3 x,
\end{equation}
where $\sigma$ is the entropy production per unit volume per unit time
and the integral is taken over the system volume.  The specific
entropy production rate in this picture is,
\begin{equation}\label{macrosig}
\sigma= -\frac{1}{T_{\rm K}^2}\bs{q} \bcdot \vnab T_{\rm K} -
\frac{1}{T_{\rm K}} \tensor{\Pi}: \vnab \vvz,
\end{equation}
where $T_{\rm K}$ is the kinetic temperature, $\bs{q}$ is the heat
conduction flux, $\tensor{\Pi}$ is the anisotropic pressure tensor,
and $\vvz$ is the mean velocity.  As usual, the kinetic temperature
is a measurement of the random kinetic energy in a small region;
$T_{\rm K}=(m/3k_{\rm B})\avps$, where $k_{\rm B}$ is Boltzmann's
constant and $v_p$ is the magnitude of the peculiar velocity. 
The heat conduction flux $\bs{q}=\qvec$ represents the peculiar kinetic
energy that is transported by peculiar velocities in the system.
Interested readers may find details of the calculation of $\sigma$ in
\citet[][\S 3.3]{dgm84}.

In order to attempt to follow the behavior of entropy from this
macroscopic viewpoint during a simulation, we form the necessary
macroscopic quantities by averaging over spatial volumes.  Simulated
systems are divided into spherical volume elements, and values for
temperature and pressure tensor components are found using particles
within the volume element.  While this procedure is straightforward,
the average number of particles per element can be rather small, even
for modest spherical grid resolutions.  For example, in a simulation
with $N=10^5$ particles broken into a spherical grid with 10 radial, 8
polar, and 8 azimuthal bins, each will contain on the order of 100
particles.  The need for gradients in quantities places some
constraint on how coarse the spherical grid can be made.  We use
simulations with $N=10^6$ particles and the aforementioned grid
resolution to begin to investigate this macroscopic picture of entropy
behavior during an evolution.  Results of these calculations are
discussed in Section~\ref{entprod}.

\section{Density and Velocity Profile Fitting}\label{fitting}

We take the GADGET simulations with $N=10^5$ as the standards for the
majority of our results.  These simulations provide a wide-range of
evolutions from which we draw broad inferences.  The other simulations
(ensemble $N=10^5$ GADGET, $N=10^6$ GADGET, GPU-NBODY-6) are focused
on testing relationships and behaviors suggested by the standard set.

\subsection{Individual $N=10^5$ GADGET Simulations}

Lynden-Bell and Maxwell-Boltzmann models can provide excellent fits to
the density and velocity profiles of the final equilibrium states of
single and hot-clumpy systems when $Q_0 \ga 0.6$.  However,
independent of the type of particle distribution (single or clumpy)
and initial density profile, systems with $Q_0 \la 0.6$ do not evolve
to states like those predicted by our maximum entropy argument.  The
density and velocity distributions of these simulated equilibria show
significant deviations from thermodynamic model expectations.  For
systems with small enough $Q_0$ and/or cold-clumpy initial
conditions, the final density distributions are best-described by a de
Vaucouleurs profile, due to its cuspy nature.  However, the specifics
of the central density cusp do not always agree with the $\delta=1/2$
value of the de Vaucouleurs profile.  We have effectively let $\delta$
vary in a few cases, finding that $1/4 \la \delta \la 3/4$.  The
thermodynamic models investigated here simply cannot reproduce the
central density cusp.  Interestingly, Plummer models provide good
descriptions of the density profiles of warm-clumpy systems
when $Q_0 \ga 0.6$.

\begin{figure}
\plotone{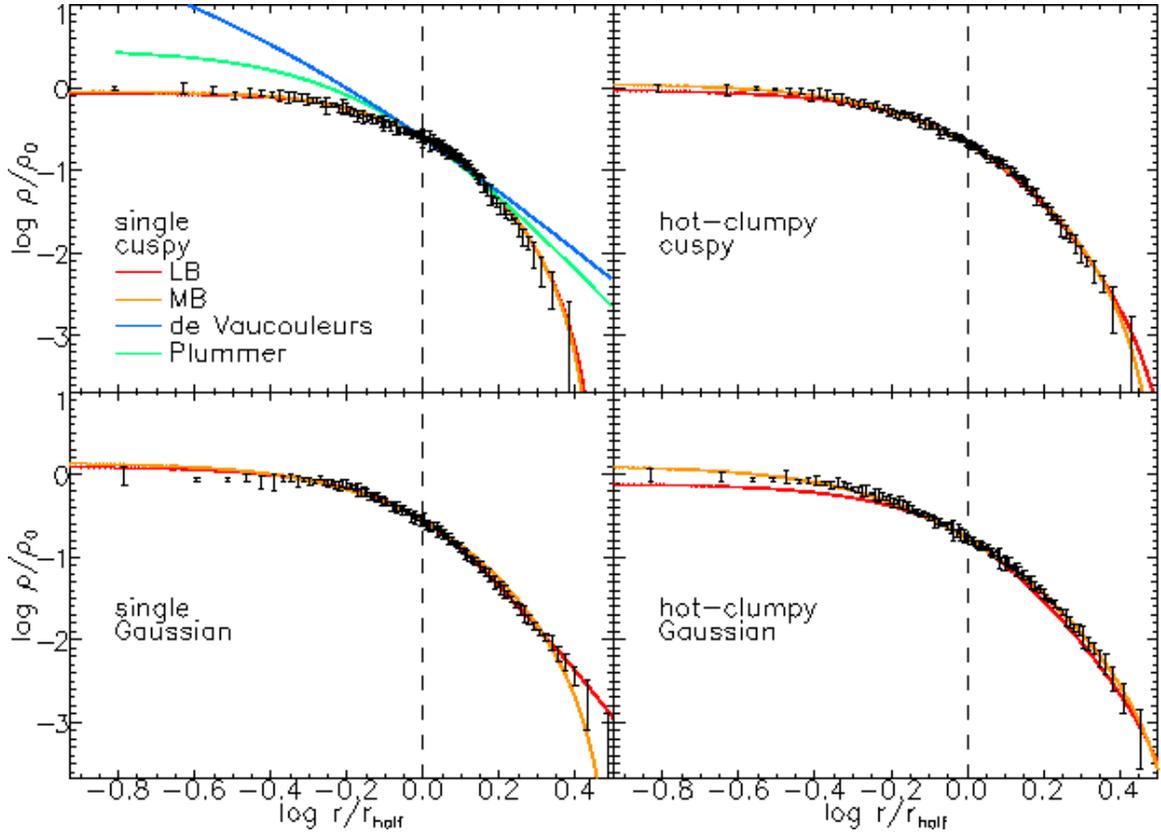}
\caption{Logarithmic density profiles for individual $N=10^5$ GADGET
simulations with $Q_0=1.0$.  The radial coordinate is scaled by the
half-mass radius (indicated by the vertical dashed line) in all
profiles.  In each panel, the initial conditions are specified and the
errorbars indicate the data values from the simulations.  The curves
show the behaviors of the various models, specified in the legend.  As
they provide poor descriptions of the simulations, the Plummer and de
Vaucouleurs models are only included in the single cuspy panel for
reference.  Lynden-Bell (LB) model fits produce density \cs\ values
smaller or comparable to those with Maxwell-Boltzmann (MB) models
(except for the hot-clumpy Gaussian simulation):
single cuspy -- $\cslb=0.635$, $\csmb=0.635$; single Gaussian --
$\cslb=0.811$, $\csmb=2.162$; hot-clumpy cuspy -- $\cslb=1.207$,
$\csmb=0.554$; hot-clumpy Gaussian -- $\cslb=5.093$, $\csmb=0.581$.
\label{den1}}
\end{figure}

\begin{figure}
\plotone{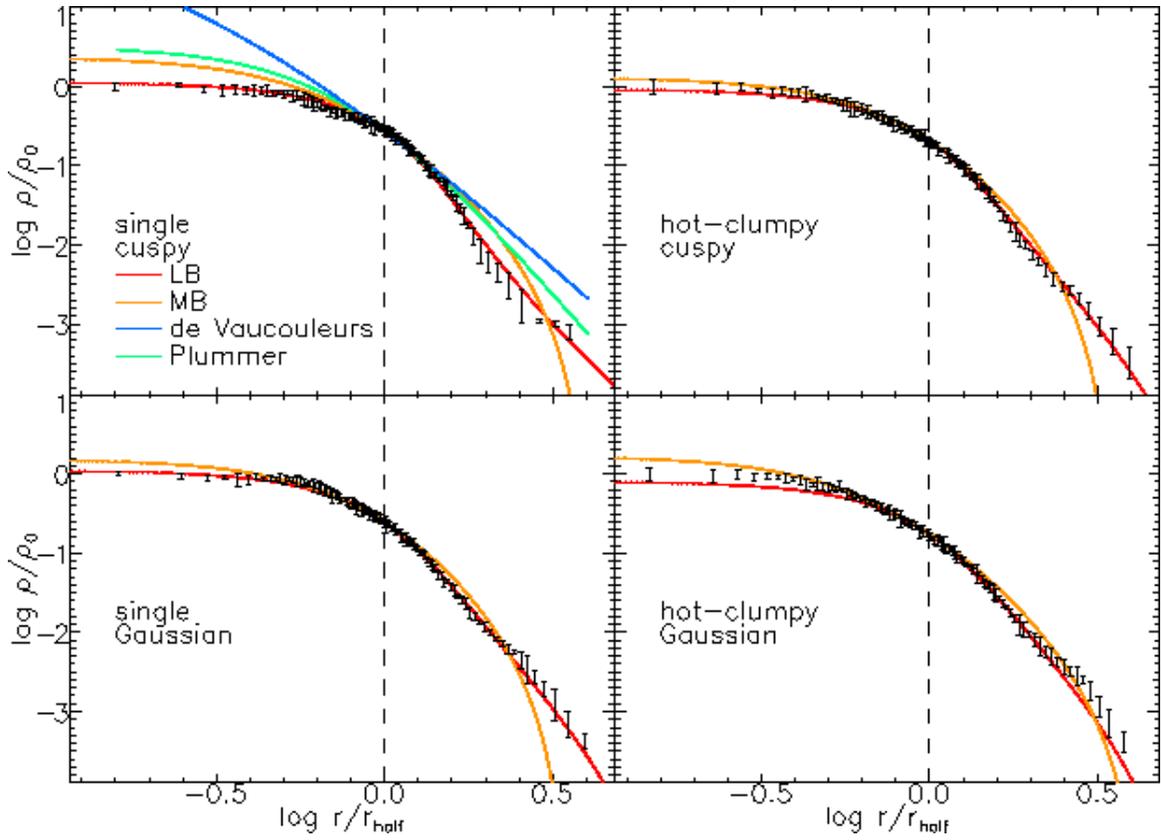}
\caption{Logarithmic density profiles for individual $N=10^5$ GADGET
simulations with $Q_0=0.7$.  The panels and linestyles are the same as
in Figure~\ref{den1}.  In these simulations, the density profiles
share a slight upward concavity in their outer profile.  In general,
the LB model fits are able to match this data behavior better than the
MB models: single cuspy -- $\cslb=1.503$, $\csmb=9.218$; single
Gaussian -- $\cslb=0.903$, $\csmb=4.847$; hot-clumpy cuspy --
$\cslb=0.777$, $\csmb=4.332$; hot-clumpy Gaussian -- $\cslb=3.204$,
$\csmb=3.360$.  
\label{den2}}
\end{figure}

Now, we turn to the simulations where there is better agreement
between the thermodynamic models and the simulations.  As
illustrations of the quality of these fits, Figures~\ref{den1} and
\ref{den2} contain plots of logarithmic density profiles for single
and hot-clumpy equilibrium systems evolved from both initial density
profiles.  In each panel, the LB and MB models are superimposed
(comparison Plummer and de Vaucouleurs profiles are also shown in the
upper-left-hand panel).  To highlight the range of density profile
behaviors, Figure~\ref{den1} contains results of simulations with
$Q_0=1.0$, while Figure~\ref{den2} shows results from evolutions with
$Q_0=0.7$.  Analogous plots for simulations with $Q_0=0.8$ and
$Q_0=0.9$ (not shown) reveal very similar model behaviors.  For single
systems, LB models produce fits superior to those from MB models in
every case.  For hot-clumpy systems, MB models perform better than LB
models for $Q_0 \ge 0.9$, with the models reversing positions for $Q_0
\le 0.8$.  Overall, LB models appear to do a better job of describing
the density distributions of the simulated equilibria.

\begin{figure}
\plotone{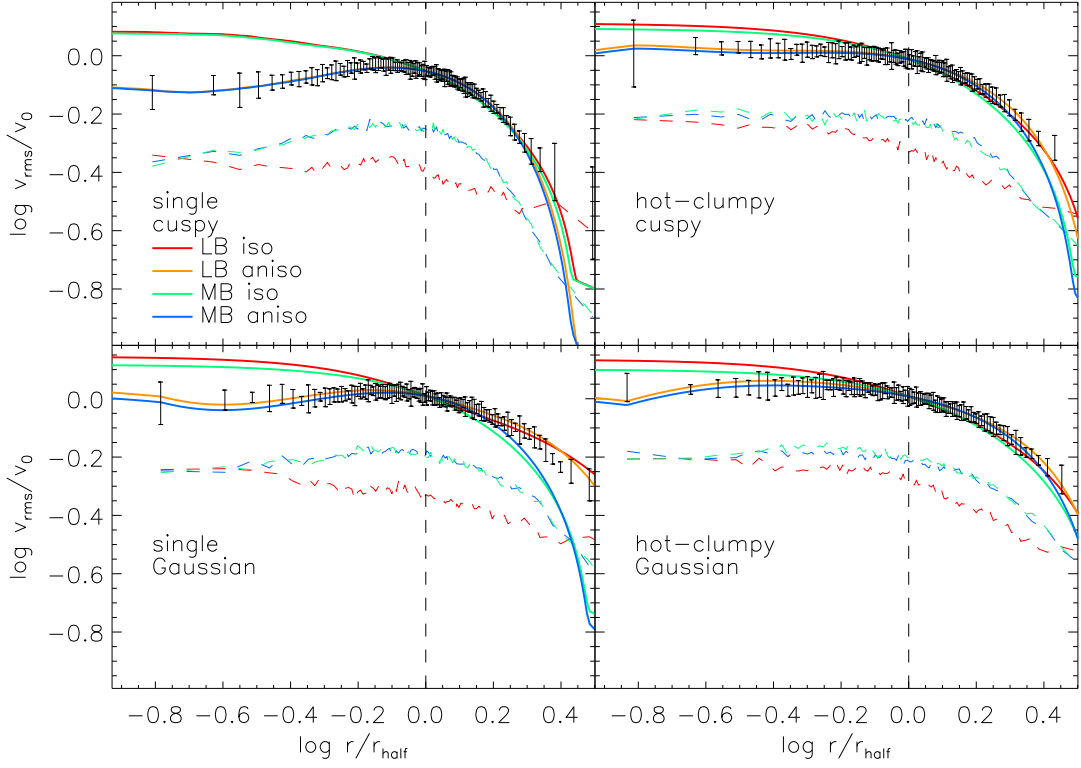}
\caption{Logarithmic \vrms\ profiles for individual $N=10^5$ GADGET
simulations with $Q_0=1.0$.  Again, the radial coordinate is scaled by
the half-mass radius (vertical dashed line).  As before, the initial
conditions are specified in each panel, and the errorbars indicate the
data values from the simulations.  The thin lines that appear below
the errorbars show the behaviors of $v_{{\rm rms},r}$ (the lowest
line) and the nearly identical $v_{{\rm rms},\theta}$ and $v_{{\rm
rms},\phi}$.  From these components, one can see the mild tangential
velocity anisotropy that exists in these simulated equilibria.  Two
sets of model curves are shown superimposed with the data.  As
indicated in the legend in the upper-left panel, there are isotropic
and anisotropic model results.  Isotropic model profiles reach nearly
constant values near the centers of systems, while anisotropic model
profiles provide better representations of the data for smaller $r$.
In general, anisotropic LB models produce the smallest velocity \cs\
values: single cuspy -- $\cslbi=1.108$, $\cslba=0.036$,
$\csmbi=1.072$, $\csmba=0.050$; single Gaussian -- $\cslbi=1.087$,
$\cslba=0.078$, $\csmbi=1.778$, $\csmba=0.832$; hot-clumpy cuspy --
$\cslbi=0.466$, $\cslba=0.016$, $\csmbi=0.500$, $\csmba=0.087$;
hot-clumpy Gaussian -- $\cslbi=0.447$, $\cslba=0.038$, $\csmbi=0.325$,
$\csmba=0.033$.  Coupled with the generally better density fits
provided by the LB models, these velocity fits suggest that the LB
models are the superior description of the phase-space distributions
of these simulations.
\label{vel1}}
\end{figure}

\begin{figure}
\plotone{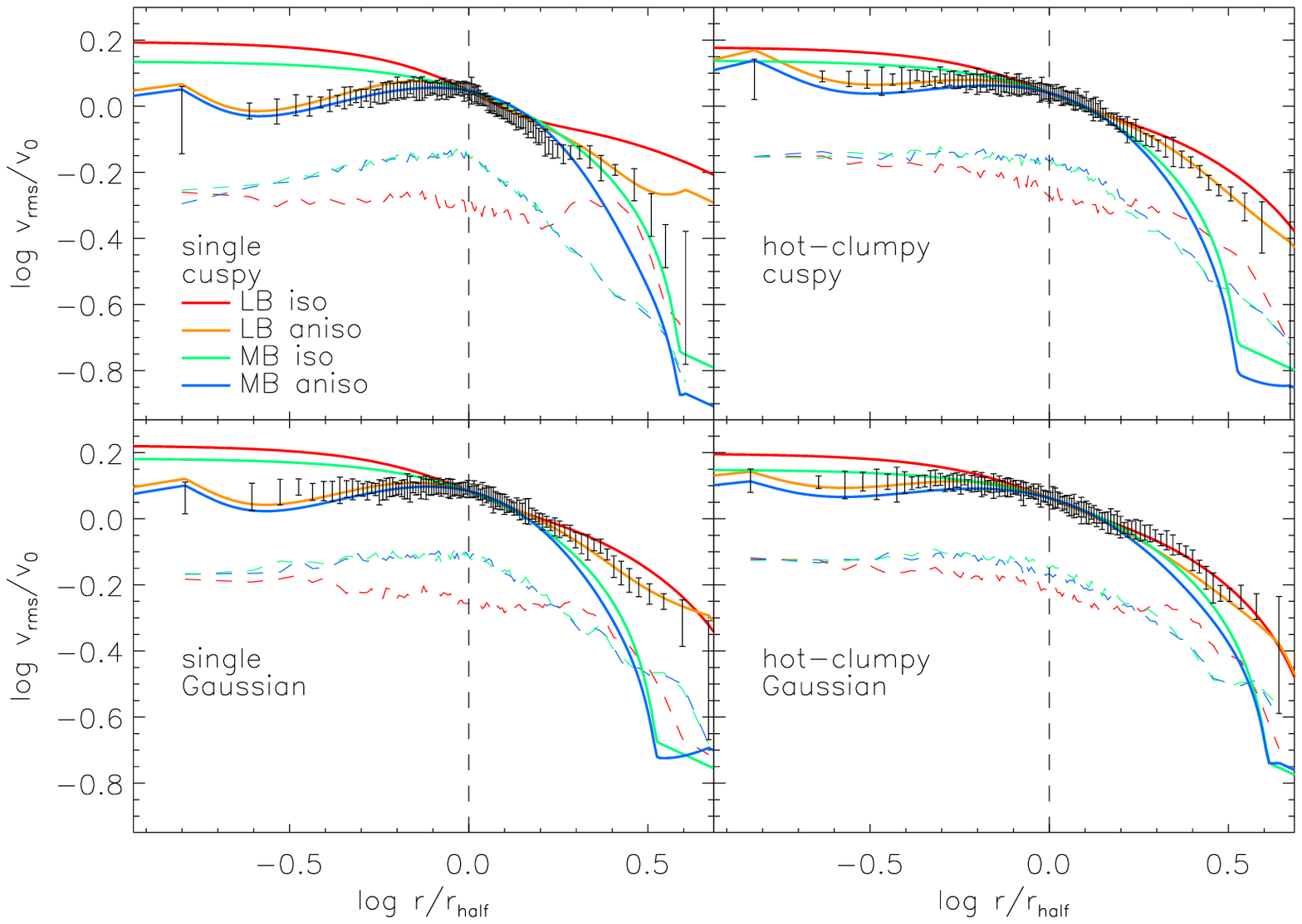}
\caption{Logarithmic \vrms\ profiles for individual $N=10^5$ GADGET
simulations with $Q_0=0.7$.  The panels here are analogous to those in
Figure~\ref{vel1}.  The ineffectiveness of isotropic models remains
evident, and the superiority of anisotropic LB models is even clearer
than in the $Q_0=1.0$ cases: single cuspy -- $\cslbi=2.231$,
$\cslba=0.214$, $\csmbi=1.024$, $\csmba=0.644$; single Gaussian --
$\cslbi=0.953$, $\cslba=0.072$, $\csmbi=1.393$, $\csmba=1.724$;
hot-clumpy cuspy -- $\cslbi=0.696$, $\cslba=0.043$, $\csmbi=1.866$,
$\csmba=3.254$; hot-clumpy Gaussian -- $\cslbi=0.324$, $\cslba=0.167$,
$\csmbi=0.546$, $\csmba=1.130$.
\label{vel2}}
\end{figure}

Corresponding plots for the \vrms\ profiles are given in
Figures~\ref{vel1} and \ref{vel2}.  In these figures, the profiles
created assuming isotropic and anisotropic velocity distributions show
quite different behaviors.  Finding $\vrms(r)$ from the Jeans
equation under the assumption that $\beta(r)=0$ results in profiles
that have flat cores.  However, single systems generically have \vrms\
profiles with non-zero slopes near their centers.  Hot-clumpy systems
are qualitatively similar to the isotropic model predictions, but are
not terribly well described by the models.  The set of thin lines
below the data points illustrate the $v_{{\rm rms,}r}$, $v_{{\rm
rms,}\theta}$, and $v_{{\rm rms,}\phi}$ behavior of the simulation.
The differences between the line shapes indicates mild tangential
velocity anisotropy.  Allowing this $\beta(r)$ profile as input to the
Jeans equation results in \vrms\ profiles that dramatically increase
the quality of the fits.  We note that the LB models tend to provide
better fits to the simulated \vrms\ profiles than those produced by
the MB models.

\subsection{Other Simulations}

\begin{figure}
\plotone{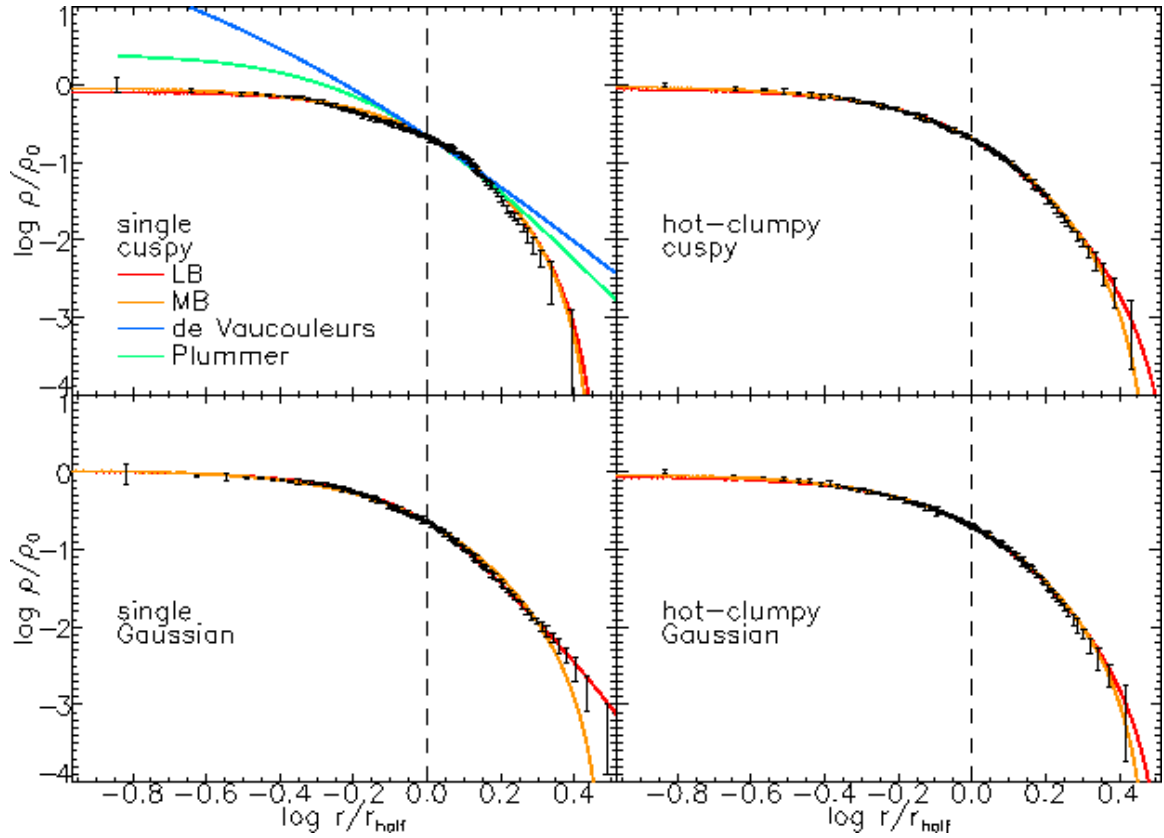}
\caption{Logarithmic density profiles for averaged $N=10^5$ GADGET
simulations with $Q_0=1.0$.  The panels are analogous to those in
Figure~\ref{den1}.  The errorbars marking the data points are smaller
and the data points show less point-to-point variation than in
Figure~\ref{den1}.  As before, the LB models tend to provide better
descriptions of the data: single cuspy -- $\cslb=4.569$,
$\csmb=4.438$; single Gaussian -- $\cslb=0.635$, $\csmb=3.182$;
hot-clumpy cuspy -- $\cslb=0.769$, $\csmb=0.652$; hot-clumpy Gaussian
-- $\cslb=1.389$, $\csmb=0.754$.
\label{aden1}}
\end{figure}

\begin{figure}
\plotone{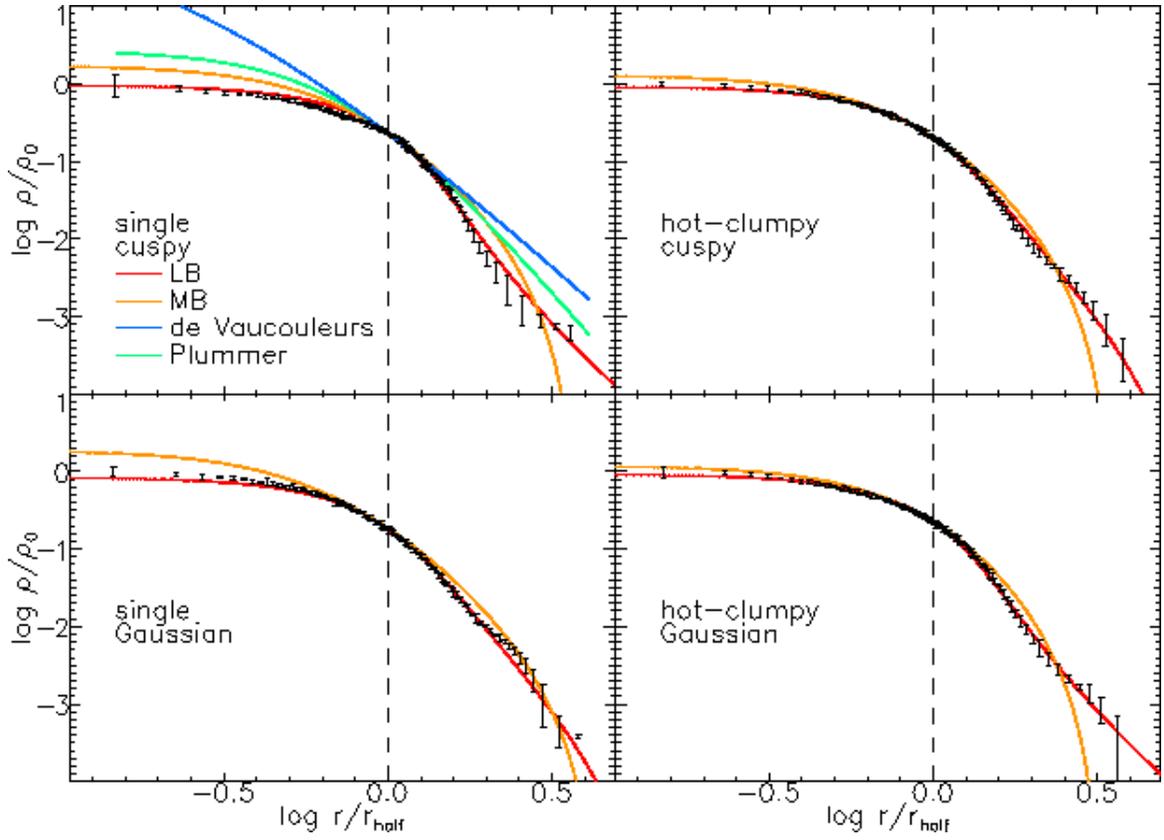}
\caption{Logarithmic density profiles for averaged $N=10^5$ GADGET
simulations with $Q_0=0.7$.  The panels are analogous to those in
Figure~\ref{den2}, and the data profiles present smoother versions of
the outer-profile concavity features seen there.  Again, the LB models
tend to describe the data behavior more completely than the MB models:
single cuspy -- $\cslb=7.658$,
$\csmb=44.058$; single Gaussian -- $\cslb=3.346$, $\csmb=28.463$;
hot-clumpy cuspy -- $\cslb=0.788$, $\csmb=7.670$; hot-clumpy Gaussian
-- $\cslb=2.215$, $\csmb=9.064$.
\label{aden2}}
\end{figure}

\begin{figure}
\plotone{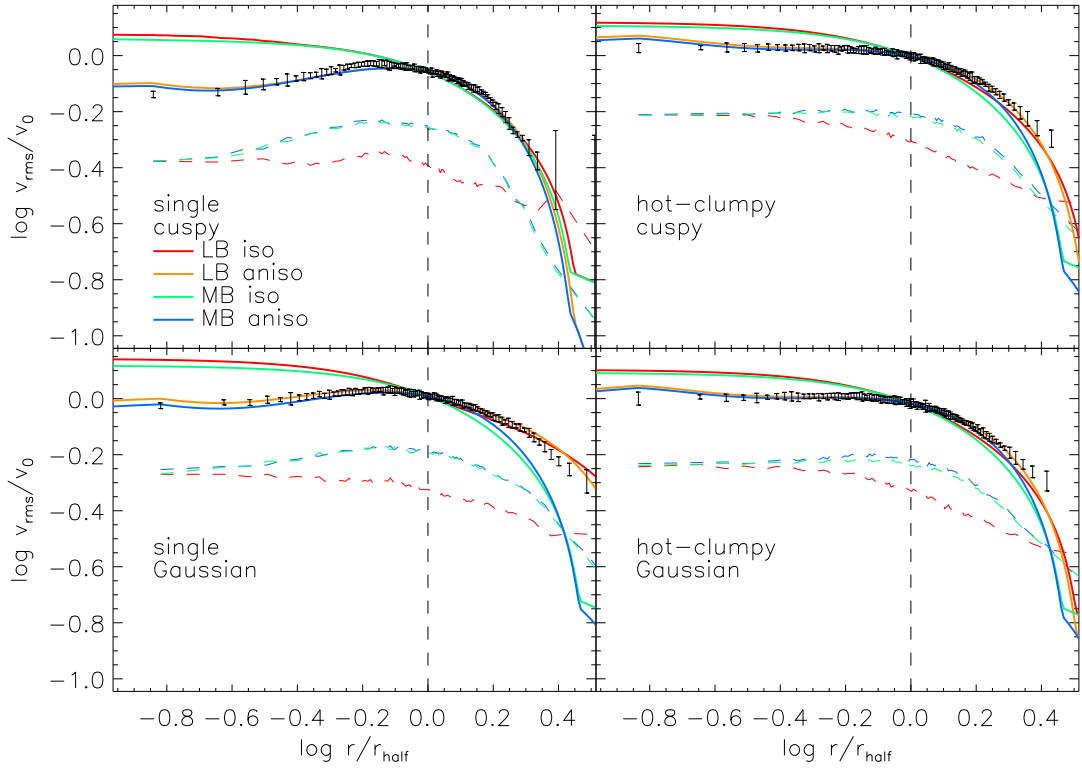}
\caption{Logarithmic \vrms\ profiles for averaged $N=10^5$ GADGET
simulations with $Q_0=1.0$.  The panels are analogous to those in
Figure~\ref{vel1}.  These smoother versions again indicate that models
incorporating the velocity anisotropy present in the simulations are
better suited to describing the data.  Of the models considered here,
the anisotropic LB models provide the best representation of the data:
single cuspy -- $\cslbi=5.866$,
$\cslba=0.370$, $\csmbi=5.629$, $\csmba=0.424$; single Gaussian --
$\cslbi=6.326$, $\cslba=0.121$, $\csmbi=9.294$, $\csmba=3.013$;
hot-clumpy cuspy -- $\cslbi=2.521$, $\cslba=0.080$, $\csmbi=3.319$,
$\csmba=0.567$; hot-clumpy Gaussian -- $\cslbi=3.137$, $\cslba=0.087$,
$\csmbi=3.531$, $\csmba=0.357$.
\label{avel1}}
\end{figure}

\begin{figure}
\plotone{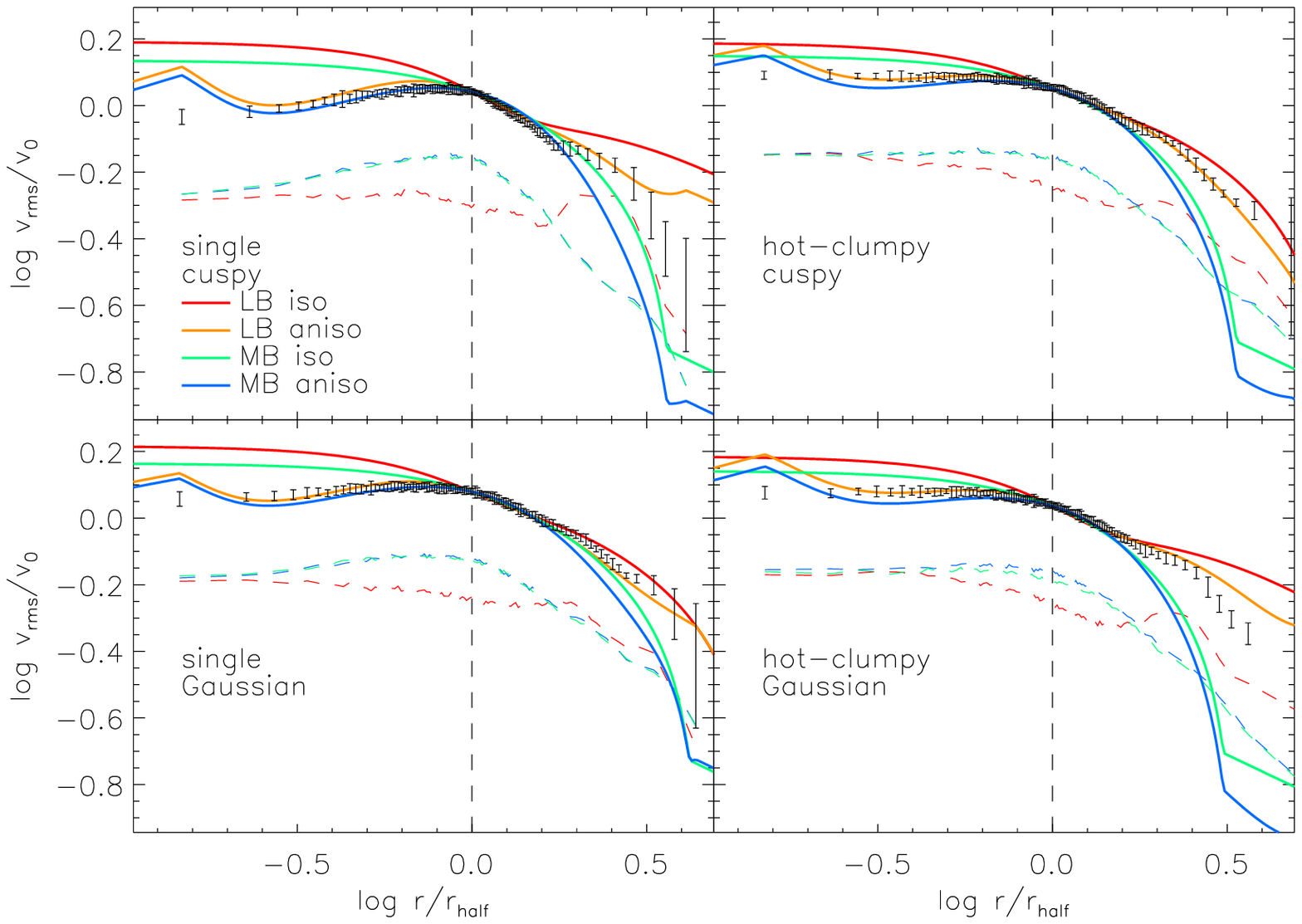}
\caption{Logarithmic \vrms\ profiles for averaged $N=10^5$ GADGET
simulations with $Q_0=0.7$.  These panels are analogous to those in
Figure~\ref{vel2}.  The smaller errorbars in these simulations result
in some poorer fits between the anisotropic LB model and the data (\eg
in the clumpy Gaussian simulation, lower-right panel):
single cuspy -- $\cslbi=8.069$,
$\cslba=0.730$, $\csmbi=3.699$, $\csmba=1.029$; single Gaussian --
$\cslbi=4.240$, $\cslba=0.341$, $\csmbi=1.945$, $\csmba=2.021$;
hot-clumpy cuspy -- $\cslbi=2.779$, $\cslba=0.214$, $\csmbi=6.954$,
$\csmba=10.852$; hot-clumpy Gaussian -- $\cslbi=2.913$, $\cslba=0.331$,
$\csmbi=2.872$, $\csmba=3.970$.
\label{avel2}}
\end{figure}

Fits to the standard simulations suggest that LB models are generally
better than MB models.  We now begin to test the robustness of this
observation using ``average'' systems formed by ensembles of
simulations with different realizations of the same initial conditions
using $N=10^5$ particles.  As with the standard simulations, GADGET has
been used to evolve the initial conditions.  Figures~\ref{aden1} and
\ref{aden2} are analogous to Figures~\ref{den1} and \ref{den2}.  The
most significant changes one notices is that the simulation profiles
are smoother and have smaller error bars.  For nearly every
simulation, LB model fits to the density distributions are superior to
those provided by the MB model (for $Q_0=1$, the two models can
produce comparable fits).  The same is true for velocity profiles.
Figures~\ref{avel1} and \ref{avel2} show the improvement that LB
models provide over MB models.  As with the standard simulations, the
inclusion of velocity anisotropy significantly improves agreement
between the model curves and the simulation results.

\begin{figure}
\plotone{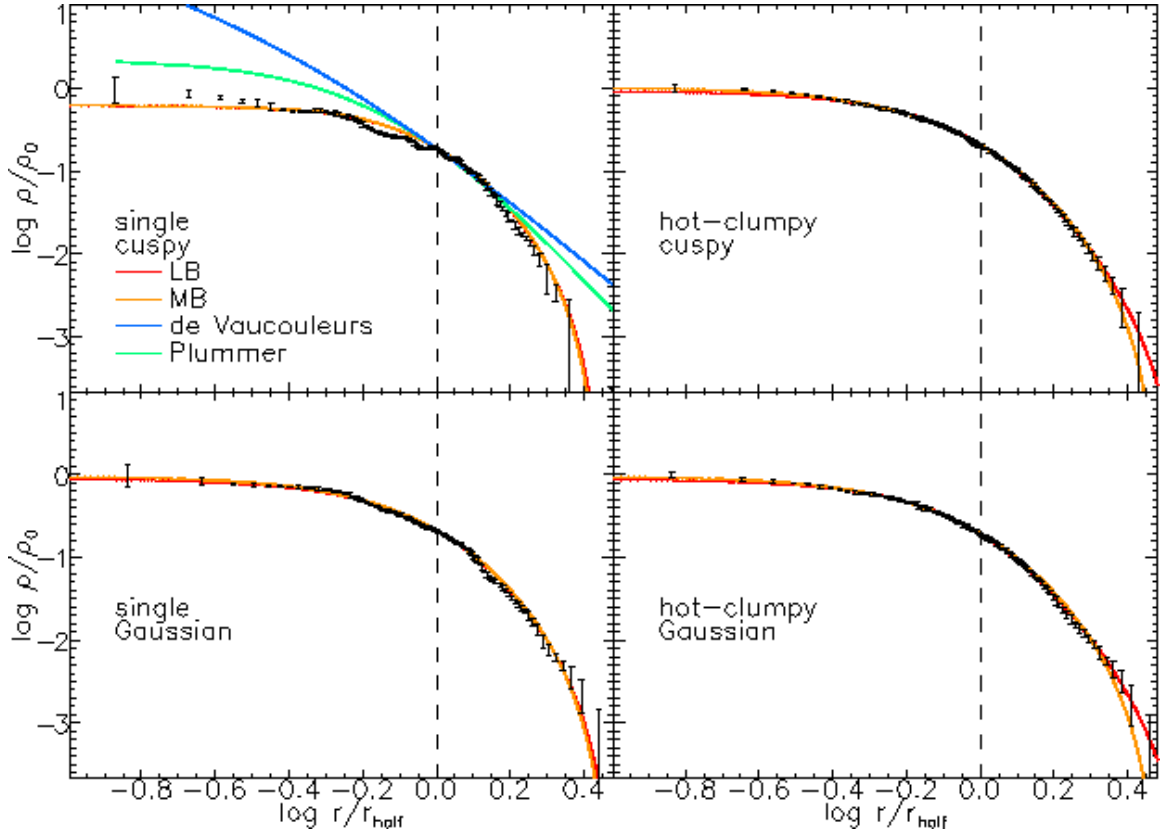}
\caption{Logarithmic density profiles for individual $N=10^6$ GADGET
simulations with $Q_0=1.0$.  These panels are analogous to those in
Figure~\ref{den1}.  The point-to-point variations seen in the single
simulations can be relatively large, leading us to question the
appropriateness of our estimated data point uncertainties.  However,
the clumpy simulation results are quite smooth, and at least on a
relative basis, the LB models describe the data better than the MB
models: single cuspy -- $\cslb=15.891$,
$\csmb=16.596$; single Gaussian -- $\cslb=6.956$, $\csmb=7.218$;
hot-clumpy cuspy -- $\cslb=0.844$, $\csmb=0.859$; hot-clumpy Gaussian
-- $\cslb=0.561$, $\csmb=1.364$.
\label{mden1}}
\end{figure}

\begin{figure}
\plotone{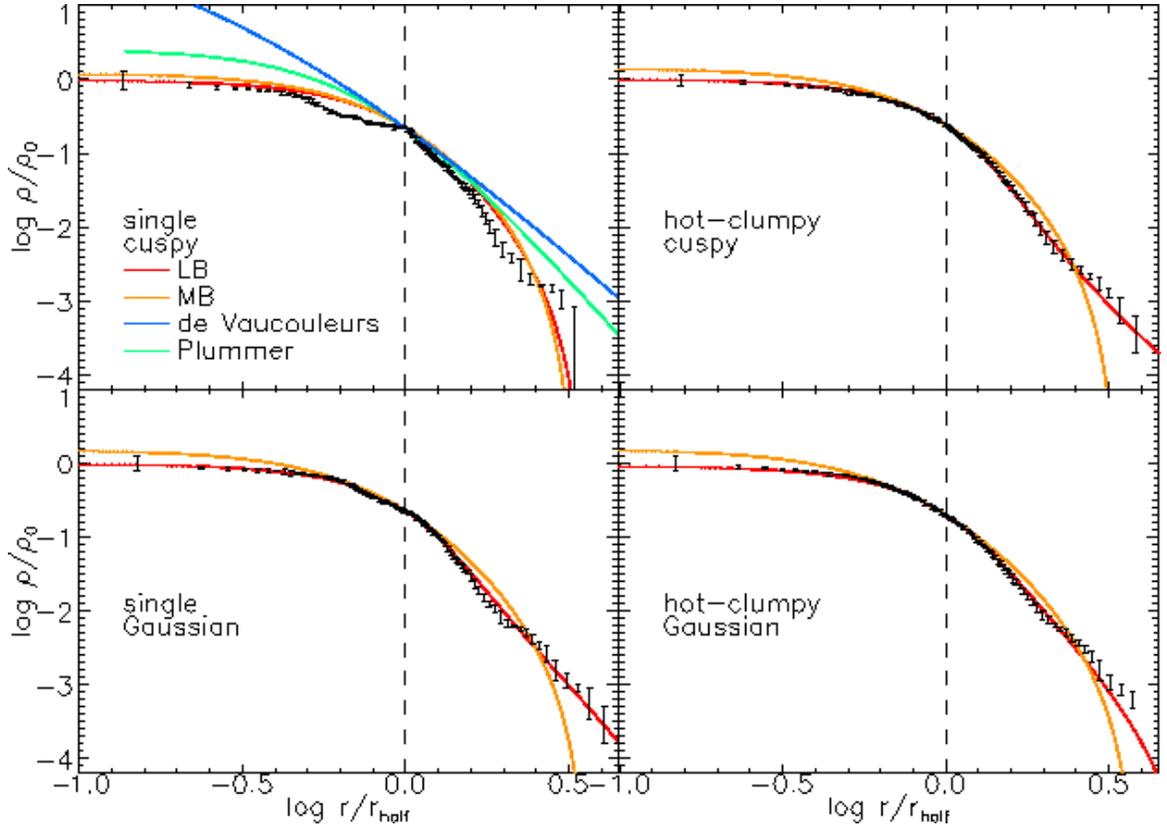}
\caption{Logarithmic density profiles for individual $N=10^6$ GADGET
simulations with $Q_0=0.7$.  These panels are analogous to those in
Figure~\ref{den2}.  Again, the data points for clumpy simulations show
a smoothness not present in the single simulations.  Except for the
single cuspy simulation, the LB models clearly provide a superior
description of the data: single cuspy -- $\cslb=81.184$,
$\csmb=105.158$; single Gaussian -- $\cslb=3.572$, $\csmb=32.259$;
hot-clumpy cuspy -- $\cslb=1.999$, $\csmb=28.178$; hot-clumpy Gaussian
-- $\cslb=2.730$, $\csmb=22.357$.
\label{mden2}}
\end{figure}

The equilibrium structures in simulations with $N=10^6$ particles
(returning to only one realization per initial condition) have also
been analyzed in a similar manner.  Figures~\ref{mden1} and
\ref{mden2} show how the LB and MB density profiles compare to the
simulation results.  The single system profiles (particularly for the
cuspy initial profile) display small-scale variations that are larger
than the uncertainty estimates we have made.  The systems are in
virial equilibrium, and the near-zero average velocity values for all
components indicate mechanical equilibrium as well.  Additional
evolutions of these initial conditions with different (smaller)
softening lengths have produced very similar outcomes.  Extending the
evolutions to longer times does reduce the variations somewhat, and we
have used the results of our longest evolutions (out to $30 T$) to
create the relevant figures.  Somewhat surprisingly, we do not see
similar structures in the profiles of the hot-clumpy simulations,
which are very smooth.  As these new features suggest that our naive
uncertainty estimates are questionable, we do not place much stock in
the actual values of \cs\ that we have found.  However, we argue that
since the LB and MB models are both being compared to the same data
with the same uncertainties, their relative \cs\ values distinguish
between which is the more appropriate model.

\begin{figure}
\plotone{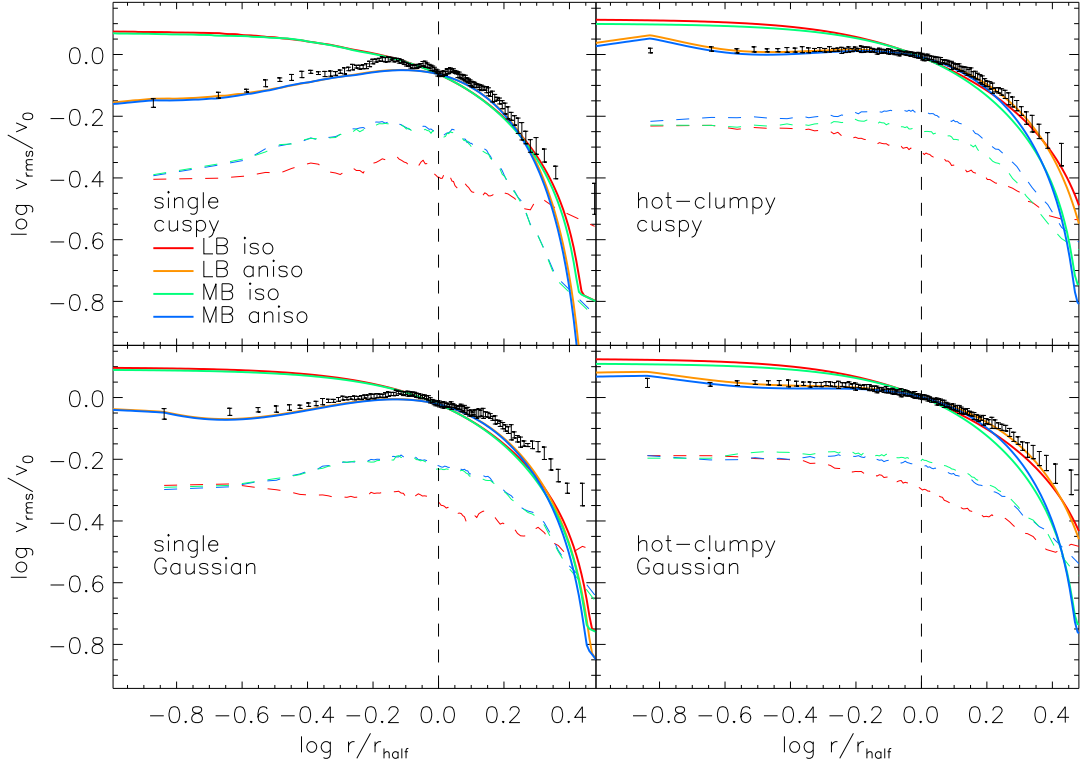}
\caption{Logarithmic \vrms\ profiles for individual $N=10^6$ GADGET
simulations with $Q_0=1.0$.  These panels are analogous to those in
Figure~\ref{vel1}.  The models with velocity isotropy continue to be
poor descriptors of the data, while the anisotropic LB models provide
the best fits to the simulation results:
single cuspy -- $\cslbi=16.044$,
$\cslba=2.962$, $\csmbi=15.723$, $\csmba=3.188$; single Gaussian --
$\cslbi=12.313$, $\cslba=3.004$, $\csmbi=12.135$, $\csmba=3.733$;
hot-clumpy cuspy -- $\cslbi=7.312$, $\cslba=0.205$, $\csmbi=6.728$,
$\csmba=0.727$; hot-clumpy Gaussian -- $\cslbi=3.574$, $\cslba=0.153$,
$\csmbi=4.383$, $\csmba=1.226$.
\label{mvel1}}
\end{figure}

\begin{figure}
\plotone{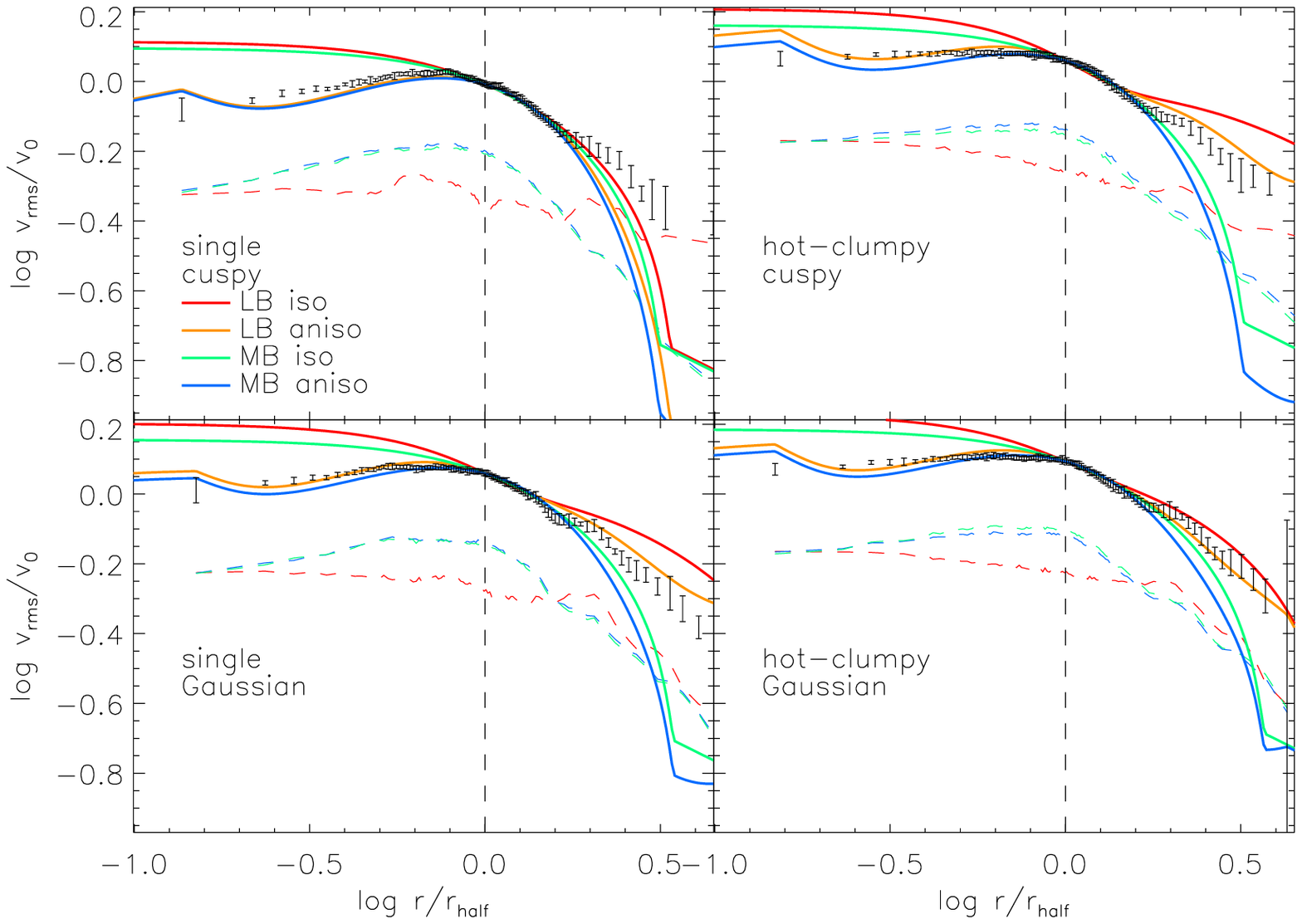}
\caption{Logarithmic \vrms\ profiles for individual $N=10^6$ GADGET
simulations with $Q_0=0.7$.  These panels are analogous to those in
Figure~\ref{vel2}.  Again, the anisotropic LB models provide the best
descriptions of the data (even though it appears to under-predict the
single cuspy simulation results for larger $r$):
single cuspy -- $\cslbi=12.735$,
$\cslba=1.824$, $\csmbi=9.868$, $\csmba=2.985$; single Gaussian --
$\cslbi=17.030$, $\cslba=0.740$, $\csmbi=7.721$, $\csmba=2.673$;
hot-clumpy cuspy -- $\cslbi=13.464$, $\cslba=0.806$, $\csmbi=5.429$,
$\csmba=2.756$; hot-clumpy Gaussian -- $\cslbi=15.484$, $\cslba=0.756$,
$\csmbi=7.722$, $\csmba=2.308$.
\label{mvel2}}
\end{figure}

Unlike the averaged simulations, the appropriateness of the LB model
is not obvious here.  In most cases, the LB and MB models provide
comparable fits to the simulation density profiles; only for the
hot-clumpy simulations with Gaussian initial density profiles is the
LB model clearly preferred.  Likewise, the differences between the LB
and MB velocity profiles are more subdued in Figures~\ref{mvel1} and
\ref{mvel2} compared to previous versions.  It also appears that there
are more significant differences between the data and the LB model
curves in these figures. 

\begin{figure}
\plotone{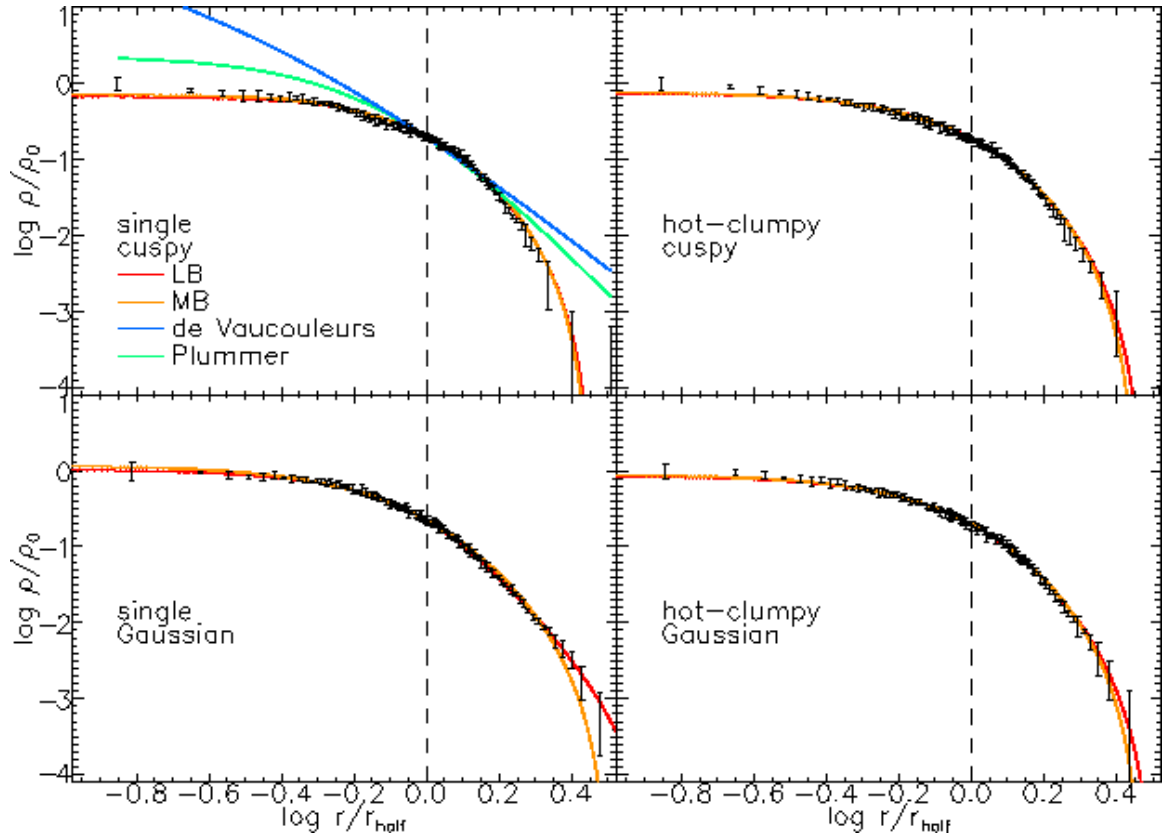}
\caption{Logarithmic density profiles for individual $N=2^{17}$
GPU-NBODY-6 simulations with $Q_0=1.0$.  These panels are analogous to
those in Figure~\ref{den1}.  Evolving systems with this non-softened
code produces equilibria with density profiles that are nearly
indistinguishable from those discussed earlier.  As with the GADGET
results, LB models provide better representations of the data:
single cuspy -- $\cslb=1.952$,
$\csmb=2.064$; single Gaussian -- $\cslb=0.390$, $\csmb=1.537$;
hot-clumpy cuspy -- $\cslb=1.096$, $\csmb=0.964$; hot-clumpy Gaussian
-- $\cslb=0.595$, $\csmb=0.584$.
\label{nb6den1}}
\end{figure}

\begin{figure}
\plotone{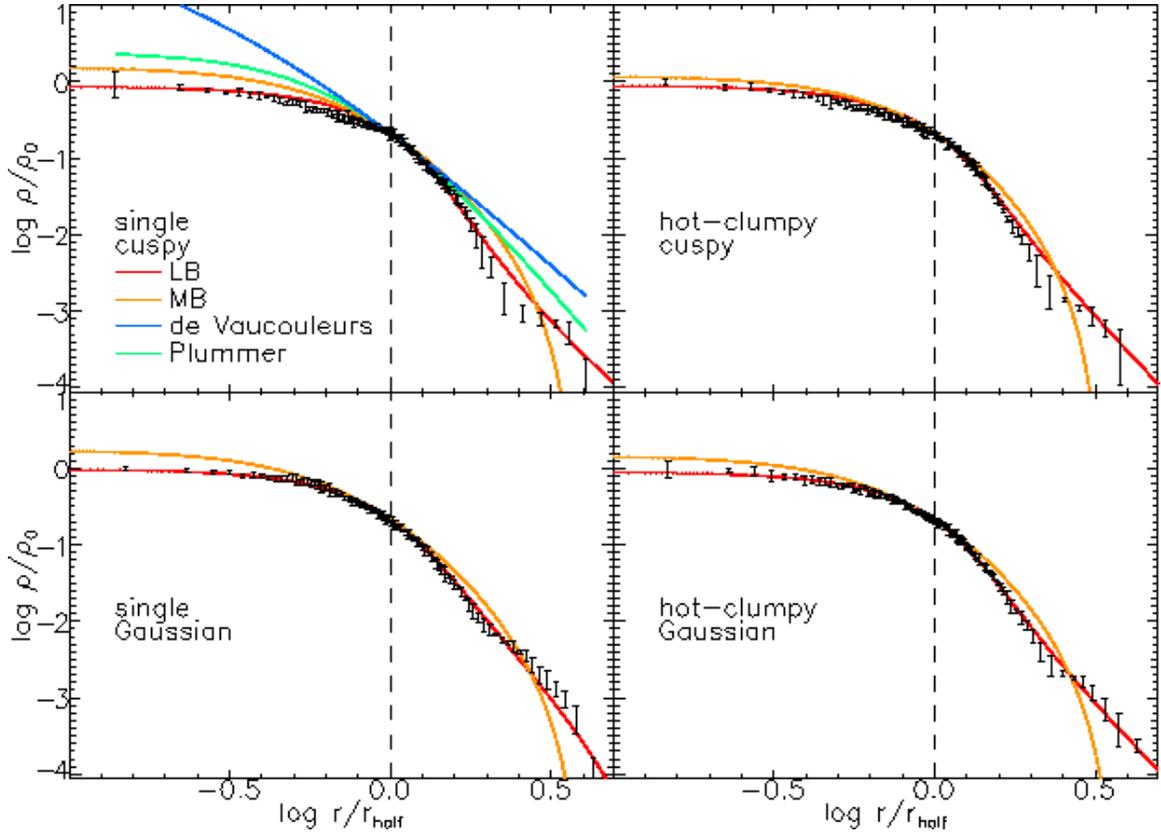}
\caption{Logarithmic density profiles for individual $N=2^{17}$
GPU-NBODY-6 simulations with $Q_0=0.7$.  These panels are analogous to
those in Figure~\ref{den2}.  The fact that the LB models can reproduce
the slight concave features in the outer regions of the profiles again
lead to them being preferred to the MB models:
single cuspy -- $\cslb=3.653$,
$\csmb=17.118$; single Gaussian -- $\cslb=1.420$, $\csmb=13.025$;
hot-clumpy cuspy -- $\cslb=3.191$, $\csmb=13.159$; hot-clumpy Gaussian
-- $\cslb=0.735$, $\csmb=8.175$.
\label{nb6den2}}
\end{figure}

\begin{figure}
\plotone{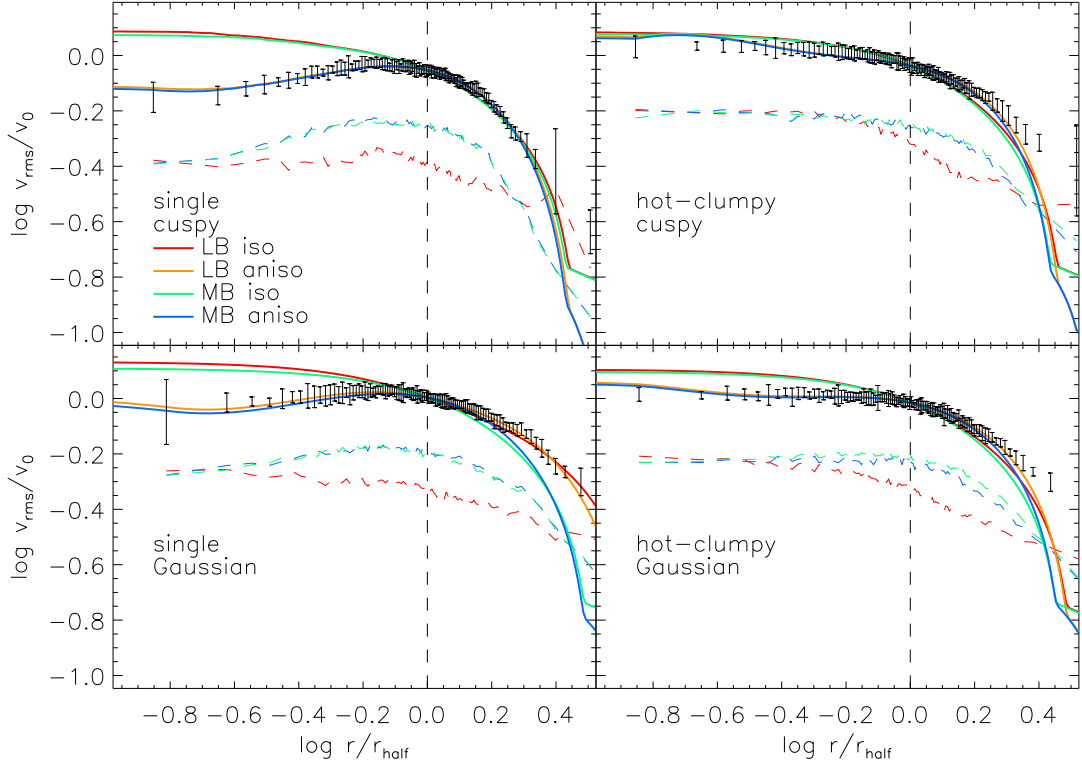}
\caption{Logarithmic \vrms\ profiles for individual $N=2^{17}$
GPU-NBODY-6 simulations with $Q_0=1.0$.  These panels are analogous to
those in Figure~\ref{vel1}.  The trend for anisotropic LB models to
provide the best description of the \vrms\ data continues.  Note that
the clumpy cuspy simulation results in a nearly isotropic velocity
distribution, and that the inner part of the profile is decently
well-described by the isotropic LB prediction:
single cuspy -- $\cslbi=1.741$,
$\cslba=0.048$, $\csmbi=1.639$, $\csmba=0.058$; single Gaussian --
$\cslbi=0.904$, $\cslba=0.032$, $\csmbi=1.734$, $\csmba=0.953$;
hot-clumpy cuspy -- $\cslbi=0.508$, $\cslba=0.194$, $\csmbi=0.711$,
$\csmba=0.310$; hot-clumpy Gaussian -- $\cslbi=1.278$, $\cslba=0.192$,
$\csmbi=1.787$, $\csmba=0.560$.
\label{nb6vel1}}
\end{figure}

\begin{figure}
\plotone{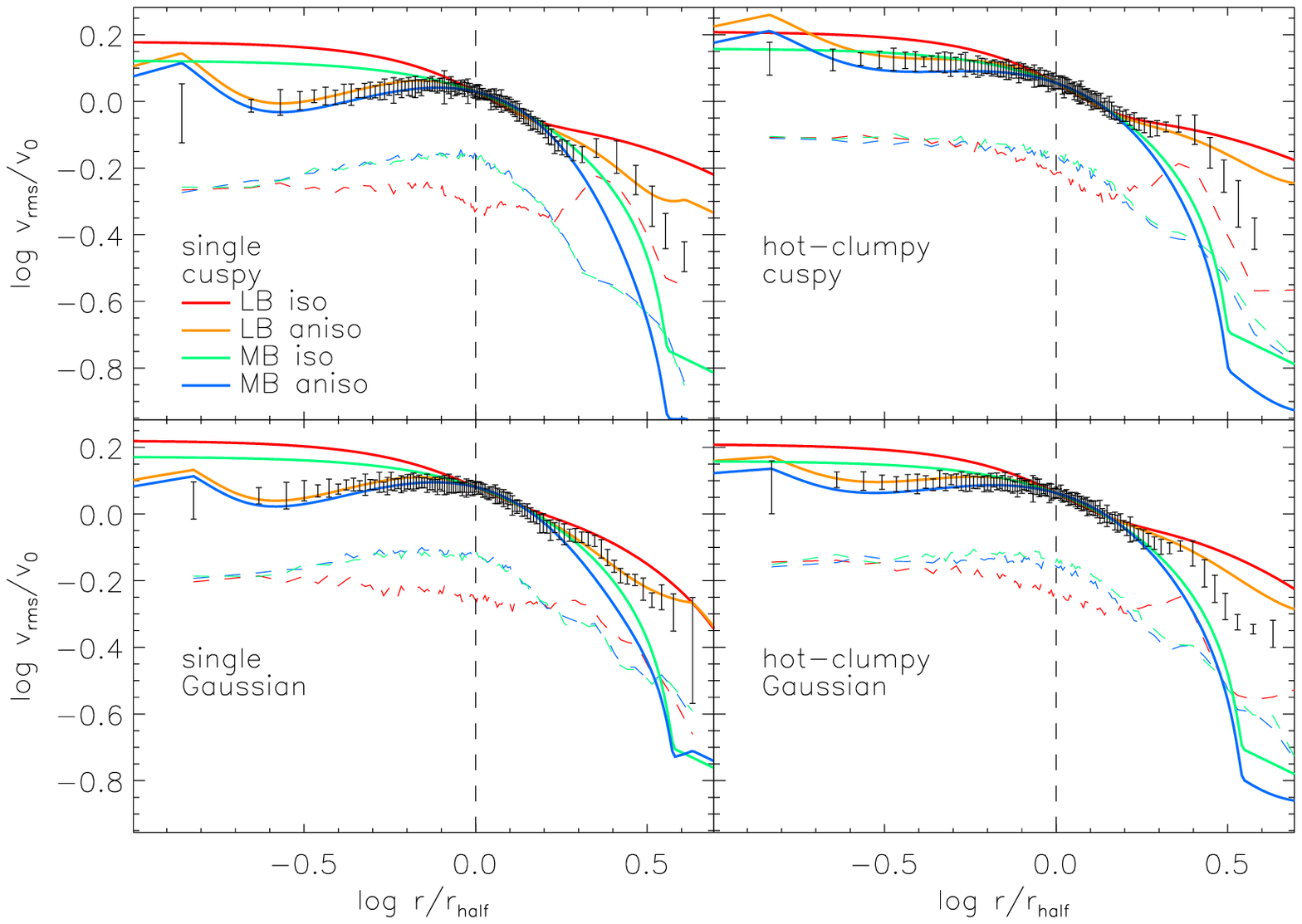}
\caption{Logarithmic \vrms\ profiles for individual $N=2^{17}$
GPU-NBODY-6 simulations with $Q_0=0.7$.  These panels are analogous to
those in Figure~\ref{vel2}.  The anisotropic LB models continue to
represent the data behavior better than the other models considered:
single cuspy -- $\cslbi=1.899$,
$\cslba=0.128$, $\csmbi=0.863$, $\csmba=0.725$; single Gaussian --
$\cslbi=1.956$, $\cslba=0.114$, $\csmbi=1.365$, $\csmba=1.772$;
hot-clumpy cuspy -- $\cslbi=0.516$, $\cslba=0.147$, $\csmbi=0.665$,
$\csmba=1.177$; hot-clumpy Gaussian -- $\cslbi=1.560$, $\cslba=0.384$,
$\csmbi=0.877$, $\csmba=1.306$.
\label{nb6vel2}}
\end{figure}

Overall, the GADGET-based simulations support the idea that the LB
model does a good job describing the density and velocity profiles of
the equilibria of collisionless systems that have undergone mild
gravitational potential relaxation.  Our final test for this idea is
to use the different evolution code GPU-NBODY-6.  Unlike GADGET,
GPU-NBODY-6 does not incorporate gravitational softening and instead
treats particles as true point masses.  While two-body encounters do
occur, the large particle number ($N > 10^5$) minimizes the global
impact of particle-particle interactions, leaving a basically
collisionless evolution.  The results of these simulations are similar
to those from the standard simulations.  In general, LB and MB models
provide comparable descriptions of the density profiles of systems
with $Q_0 \ge 0.9$, but LB models are superior for $Q_0 \le 0.8$ (see
Figures~\ref{nb6den1} and \ref{nb6den2}).  Figures~\ref{nb6vel1} and
\ref{nb6vel2} show that LB models tend to provide better fits to
\vrms\ profiles produced by these simulations.

\section{Entropy Production}\label{entprod}

\begin{figure}
\plottwo{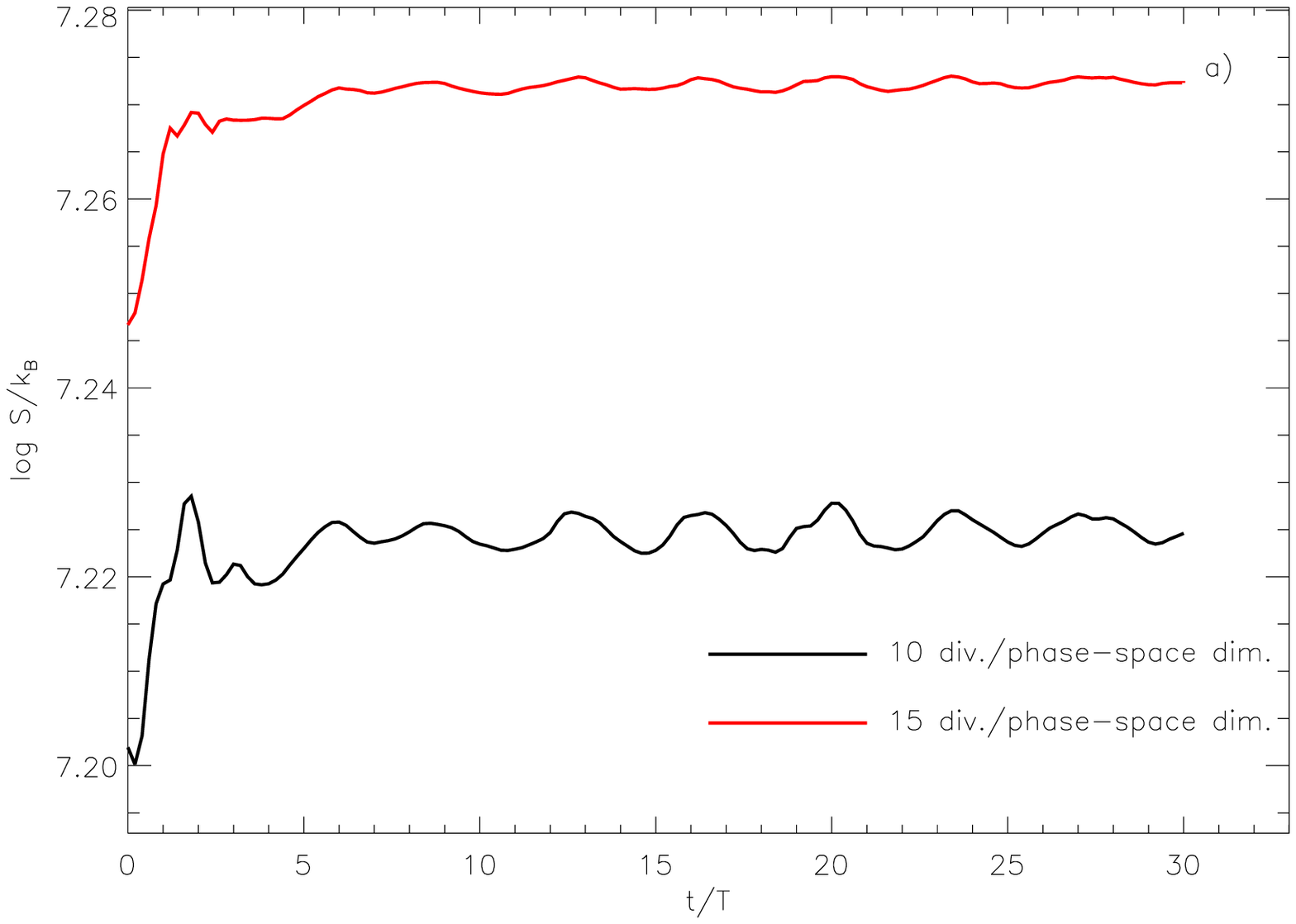}{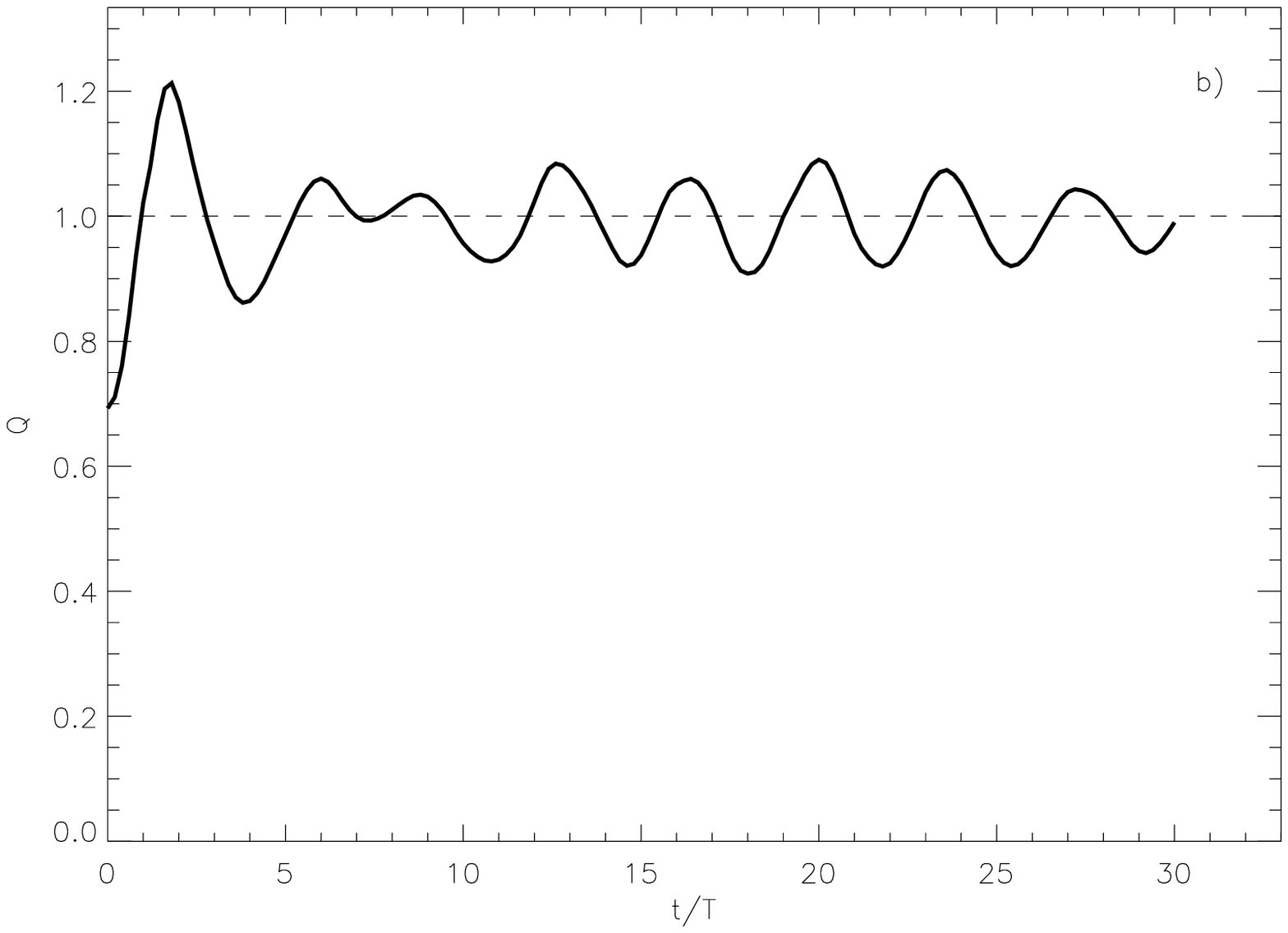}
\caption{(a) Lynden-Bell entropy versus time in the individual single,
cuspy $N=10^6$ GADGET simulation with $Q_0=0.7$.  The entropy is
calculated from the phase-space macro-cell occupation values
(Equation~\ref{slb}), with a value of $\nu_{\rm LB}=10^4$.  Two
different macro-cell volumes (indicated in the legend) have been used
to create comparison curves.  Adopting a smaller macro-cell volume
increases the zero point of the curve and reduces the small-scale
variations in $S_{\rm LB}$ but leaves the overall behavior unchanged.
(b) The virial ratio $Q$ as a function of time for the same
simulation.  The initial increase in $S_{\rm LB}$ appears to coincide
with the initial growth in $Q$.
\label{svst}}
\end{figure}

As mentioned in section~\ref{micro}, the microscopic calculation of the
entropy has been carried out for the $N=10^6$ GADGET simulations.  We
have created two different macro-cell grids; one with 10 macro-cell
divisions per phase-space dimension ($M=10^6$), and one with 15
divisions ($M \approx 10^7$).  We show a representative pair of
$S_{\rm LB}(t)$ curves in Figure~\ref{svst}a.  The different curves
correspond to the grid choices indicated.  These are derived from a
single, cuspy $Q_0=0.7$ system, but similar behavior is seen in
simulations with other $Q_0$, Gaussian density profiles, and clumpy
particle distributions.  Most importantly, it is clear that the
entropy rises from an initial value and rapidly approaches a maximum,
steady-state value.  The most rapid increase (a roughly 5\% change) in
entropy occurs during the first couple of initial crossing times.  It
seems natural that the most rapid growth in $S_{\rm LB}$ occurs during
the period when the strongest gravitational potential relaxation
(biggest variations in potential) occur.  For comparison, the
variation in the virial ratio $Q=2K/|W|$ as a function of time in the
same simulation is shown in Figure~\ref{svst}b. 

We also note that the finer macro-cell grid produces smaller
variations in the value of $S_{\rm LB}$, a trend that also occurs as
$Q_0 \rightarrow 1$.  With the increasing smoothness of the curves
with higher $Q_0$, one can also discern that there is slower growth of
$S_{\rm LB}$ occurring over tens of crossing times.  We speculate
that this slower growth is attributable to the phase mixing that
continues after the initial potential variations decrease.

\begin{figure}
\plotone{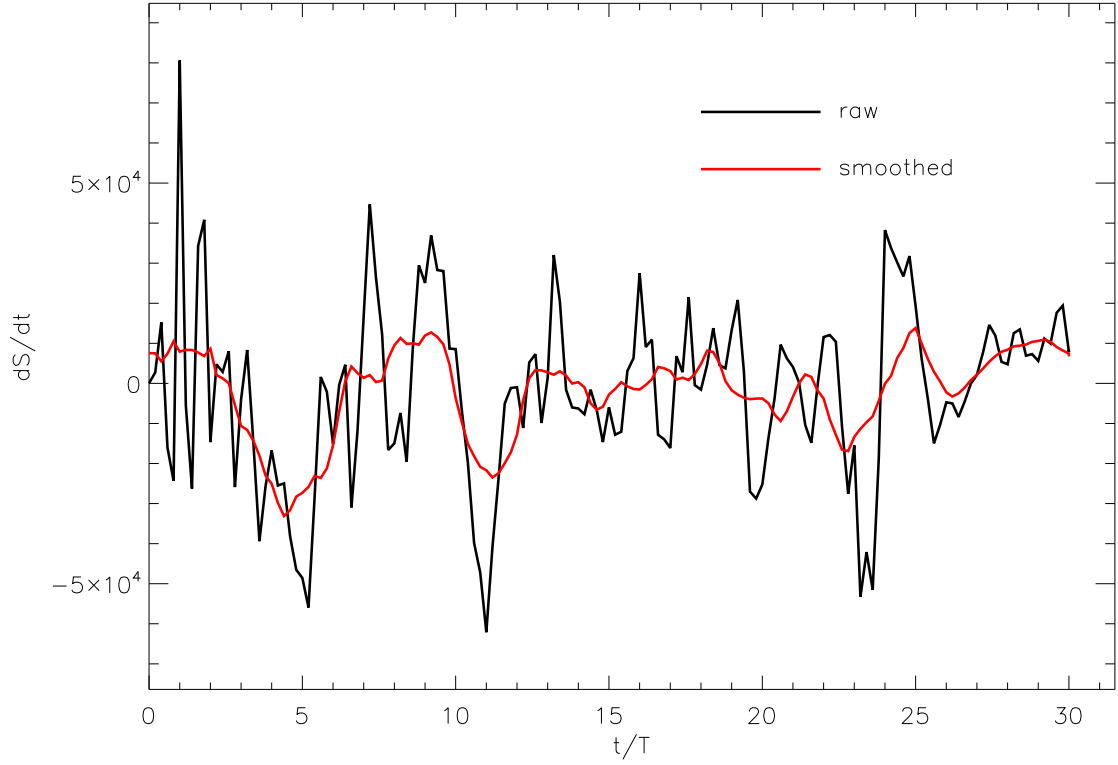}
\caption{The macroscopic entropy production rate versus time in the
individual single, cuspy $N=10^6$ GADGET simulation with $Q_0=0.7$.
The production rate is calculated by determining gradients in the
kinetic temperature and mean velocity field (Equation~\ref{macrosig}).
The rather coarse grid used to determine the gradients may play an
important role in the essentially null result shown.  If the
macroscopic entropy production could be properly calculated, one would
expect a rather large positive spike in the interval $0 \le t/T \la 5$.
\label{dsdt}}
\end{figure}

While the microscopic entropy picture dovetails nicely with the
overall scenario of entropy production in collisionless systems, the
macroscopic picture results are not so clear.  Figure~\ref{dsdt}
illustrates the behavior of Equation~\ref{s4} for the single, cuspy
$Q_0=0.7$ simulation.  Given the microscopic results, one would expect
a rather tall, positive spike to appear in the early part of the
evolution, followed by a decline towards zero.  Instead, the very
noisy curve seems to oscillate about zero.  A boxcar smoothing filter
applied to the raw values makes the oscillation more plain.  Again,
this same behavior is seen across the various $N=10^6$ GADGET
simulations.  Given the reliance on derivative information required by
this approach, the noise present in the values is not surprising.  It
is possible that our particle number and grid resolutions are simply
not high enough to capture the true behavior of the entropy creation
term.  Preliminary testing with higher grid resolution did not produce
appreciably different results.  However, there could also be a fault
in the assumption of local thermodynamic equilibrium that underlies
the derivation of Equation~\ref{macrosig}.

\section{Conclusions}\label{conclude}

We have evolved sets of $N$-body initial conditions to determine if
the thermodynamically-motivated Lynden-Bell or Maxwell-Boltzmann
distribution functions can describe the equilibrium distributions of
particle locations and velocities.  Different evolution codes and
numerical parameters have been adopted to reduce the likelihood of
spurious findings.  These simulation results have also
been used to investigate the entropy behavior of these systems.

Initially ``cold'' systems evolve in expected fashion, forming
equilibria that show cuspy central density profiles and outer density
profiles that match the de Vaucouleurs form, not the thermodynamic
models.  However, a subset of our simulations can be well-described by
distribution functions that maximize entropy.  Initially ``hot''
systems form equilibria with density profiles that are very similar to
those produced by the Lynden-Bell distribution function.
Unfortunately, the lack of velocity isotropy in these simulations
seems to preclude the ability of the LB distribution functions to
predict \vrms\ profiles.  However, accounting for the mild tangential
anisotropy produces extremely good descriptions of simulation velocity
results.  Previous simulations involving mild velocity anisotropy
indicate very similar global behavior to fully isotropic systems
\citep[\eg][]{ma85}.  We argue that the isotropic behavior is of
central importance, with velocity anisotropy determining higher-order
corrections to density and \vrms\ profiles \citep[\eg][]{m85}.  To be
clear, introducing velocity anisotropy outside the distribution
function, as we have done, means that such models are not
self-consistent.  The anisotropic \vrms\ profiles are not predictions
of the maximum entropy argument, which assumes only constant system
mass and energy.  Anisotropic models would require entropy
maximization including at least one other constraint, along the lines
of \citep{tb05}.  Given these caveats, our results suggest that there
are conditions under which a collisionless self-gravitating system
will evolve to a maximum entropy state.  This conclusion is
strengthened by the fact that direct calculation of the entropy (using
a phase-space occupation approach) also shows a maximum value being
attained.  An alternative construction of the entropy behavior is
inconclusive, presumably due to resolution effects.

There remain some unresolved issues with the idea of these
equilibria representing maximum entropy states.  One is that maximized
entropy is normally taken to imply thermodynamic equilibrium, but
these simulated systems clearly have kinetic temperature gradients.
Is it possible that in self-gravitating systems, these two conditions
are not equivalent?  Another set of questions revolve around the role
of the $\nu$ parameter.  This value represents the number of
micro-cells that occupy any macro-cell, but it does not have an ab
initio value.  In the direct calculation, it must be larger than the
largest macro-cell occupation number $n_i$ in order for the
entropy to be well-defined.  In fitting density profiles, we have
placed no such restriction on its value.  For the simulations focused
on in this work, we have found $5 \la \nu \la 2000$.  In general,
higher $Q_0$ values couple to lower $\nu$ values.  On the other hand,
the slight positive concavity seen in the outer density profiles of
$Q_0=0.7$ simulations is reproduced by the models only when $\nu \ga
1000$.  Unsurprisingly, there is no simple answer underlying the
evolution of collisionless systems, but our results suggest there are
some new questions to ask.

\acknowledgments
The authors gratefully acknowledge support from the Wisconsin Space
Grant Consortium, through the Undergraduate Research Fellowship and
Research Infrastructure programs.  We thank Lance Hildebrand for help
with the macroscopic entropy calculations.  Thanks also to the
anonymous referee for numerous helpful suggestions.

\end{document}